\documentclass[iop,revtex4,numberedappendix,appendixfloats]{emulateapj}

\usepackage{amsmath}
\usepackage{mathrsfs }
\usepackage{lscape}

\usepackage{grffile}
\usepackage{subfigure}

\usepackage{hyperref}

\newcommand{\vect}[1]{\boldsymbol{#1}}
\newcommand*\diff{\mathop{}\!\mathrm{d}}
\newcommand*\Diff[1]{\mathop{}\!\mathrm{d^#1}}
\newcommand{\pdf}{\ensuremath{pdf}}
\newcommand{\pmodel}{\ensuremath{p_M}}
\newcommand{\MAP}{MAP}
\newcommand{\MAPs}{MAPs}
\newcommand{\RM}{{\sl RoadMapping}}
\makeatletter
\newcommand{\testlabel}[2]{%
 \protected@write \@auxout {}{\string \newlabel {#1}{{#2}{\thepage}{#2}{#1}{}} }%
 \hypertarget{#1}{#2}
}
\makeatother

\shorttitle{Action-based Dynamical Modelling for the Milky Way Disk}
\shortauthors{Trick et al.}

\begin{document}

\title{Action-based Dynamical Modelling for the Milky Way Disk\\}

\author{Wilma H. Trick\altaffilmark{1,2}, Jo Bovy\altaffilmark{3}, and Hans-Walter Rix\altaffilmark{1}}

\altaffiltext{1}{Max-Planck-Institut f\"ur Astronomie, K\"onigstuhl 17, D-69117 Heidelberg, Germany}
\altaffiltext{2}{Correspondence should be addressed to trick@mpia.de.}
\altaffiltext{3}{Department of Astronomy and Astrophysics, University of Toronto, 50 St. George Street, Toronto, ON, M5S 3H4, Canada}


\begin{abstract}
We present \RM{}, a full-likelihood dynamical modelling machinery that aims to recover the Milky Way's (MW) gravitational potential from large samples of stars in the Galactic disk. \RM{} models the observed positions and velocities of stars with a parametrized, three-integral distribution function (DF) in a parametrized axisymmetric potential. We investigate through differential test cases with idealized mock data how the breakdown of model assumptions and data properties affect constraints on the potential and DF. Our key results are: (i) If the MW's true potential is not included in the assumed model potential family, we can---in the axisymmetric case---still find a robust estimate for the potential, with only $\lesssim10\%$ difference in surface density within $|z| \leq1.1~\text{kpc}$ inside the observed volume. (ii) Modest systematic differences between the true and model DF are inconsequential. E.g, when binning stars to define sub-populations with simple DFs, binning errors do not affect the modelling as long as the DF parameters of neighbouring bins differ by $<20\%$. In addition, \RM{} ensures unbiased potential estimates for either (iii) small misjudgements of the spatial selection function (i.e., $\lesssim15\%$ at the survey volume's edge), (iv) if distances are known to within $10\%$, or (v) if proper motion uncertainties are known within $10\%$ or are smaller than $\delta \mu \lesssim 1~\text{mas yr}^{-1}$. Challenges are the rapidly increasing computational costs for large sample sizes. Overall, \RM{} is well suited to making precise new measurements of the MW's potential with data from the upcoming Gaia releases.
\end{abstract}


\section{Introduction} \label{sec:intro}

Through dynamical modelling we can infer the Milky Way's (MW) gravitational potential from stellar motions (\citealt{2008gady.book.....B,2011Prama..77...39B,2013A&ARv..21...61R}). Observational information on the 6D phase-space coordinates of stars is currently growing at a rapid pace, and will be taken to a whole new level in quantity and precision by the upcoming data from the Gaia mission \citep{2001A&A...369..339P}. Yet, rigorous and practical modelling tools that turn position-velocity data of individual stars into constraints both on the gravitational potential and on the distribution function (DF) of stellar orbits are scarce \citep{2013A&ARv..21...61R}.

The Galactic gravitational potential is fundamental for understanding the MW's dark matter and baryonic structure \citep{2012sf2a.conf...15F,2013A&ARv..21...61R,2013PhR...531....1S,2014JPhG...41f3101R} and the stellar-population-dependent orbit DF is a basic constraint on the Galaxy's formation history \citep{2013NewAR..57...29B,2015MNRAS.449.3479S}.

There is a variety of practical approaches to dynamical modelling of discrete collisionless tracers, such as the stars in the MW, e.g., Jeans modelling (\citealt{1989MNRAS.239..605K}; \citealt{2012ApJ...756...89B}; \citealt{2012MNRAS.425.1445G}; \citealt{2013ApJ...772..108Z}; \citealt{2015MNRAS.452..956B}), action-based DF modelling (with parametric DFs: \citealt{2013ApJ...779..115B}; \citealt{2014MNRAS.445.3133P}; \citealt{2015MNRAS.449.3479S}; \citealt{2016MNRAS.tmp..817D}; with marginalization over non-parametric DFs: \citealt{2014MNRAS.437.2230M}), torus modelling (\citealt{2008MNRAS.390..429M}; \citealt{2012MNRAS.419.2251M}; \citealt{2013MNRAS.433.1411M}), or made-to-measure modelling (\citealt{1996MNRAS.282..223S}; \citealt{2007MNRAS.376...71D}; \citealt{2014MNRAS.443.2112H}). Most of them---explicitly or implicitly---describe the stellar distribution through a DF. Not all of them avoid binning to exploit the full discrete information content of the data.

Recently, \citet{2012MNRAS.426.1328B} and \citet{2013ApJ...779..115B} proposed constraining the MW's gravitational potential by combining parametrized axisymmetric potential models with DFs that are simple analytic functions of the three orbital actions to model discrete data.

\citet{2013ApJ...779..115B} (BR13 hereafter) put this in practice by implementing a rigorous modelling approach for so-called mono-abundance populations (\MAPs{}), i.e, sub-sets of stars with similar $[\mathrm{Fe}/\mathrm{H}]$ and $[\alpha/\mathrm{Fe}]$ within the Galactic disk, which seem to follow simple DFs \citep{2012ApJ...751..131B,2012ApJ...755..115B,2012ApJ...753..148B}. Given an assumed (axisymmetric) model for the Galactic potential and action-based DF \citep{2010MNRAS.401.2318B,2011MNRAS.413.1889B,2013MNRAS.434..652T} they calculated the likelihood of the observed ($\vec{x},\vec{v}$) for each \MAP{}, using SEGUE G-dwarf stars \citep{2009AJ....137.4377Y}. They also accounted for the complex, but known selection function of the kinematic tracers \citep{2012ApJ...753..148B}. For each \MAP{} the modelling resulted in an independent estimate of the same gravitational potential. Taken as an ensemble, they constrained the disk surface density over a wide range of radii ($\sim 4-9~\text{kpc}$), and powerfully constrained the disk mass scale length and the stellar-disk-to-dark-matter ratio at the Solar radius. 

BR13 made however a number of quite severe and idealizing assumptions about the potential, the DF and the knowledge of observational effects. These idealizations could plausibly translate into systematic errors on the inferred potential, well above the formal error bars of the upcoming surveys with their wealth and quality of data.

In this work we present \RM{} (``\textsc{R}ecovery of the \textsc{O}rbit \textsc{A}ction \textsc{D}istribution of \textsc{M}ono-\textsc{A}bundance \textsc{P}opulations and \textsc{P}otential \textsc{IN}ference for our \textsc{G}alaxy'')---an improved, refined, flexible, robust and well-tested version of the original dynamical modelling machinery by BR13. Our goal is to explore which of the assumptions BR13 made and which other aspects of data, model and machinery limit \RM{}'s recovery of the true gravitational potential.

We investigate the following aspects of the \RM{} machinery that become especially important for a large number of stars: (i) Numerical inaccuracies must not be an important source of systematics (Section \ref{sec:likelihood_normalisation}). (ii) As parameter estimates become much more precise, we need more flexibility in the potential and DF model and efficient strategies to find the best fit parameters. The improvements made in \RM{} as compared to the machinery used in BR13 are presented in Section \ref{sec:fitting}. (iii) \RM{} should be an unbiased estimator (Section \ref{sec:largedata}).

We also explore how different aspects of the observational experiment design impact the parameter recovery: (i) We consider the importance of the survey volume geometry, size, shape and position within the MW to constrain the potential (Section \ref{sec:results_obsvolume}). (ii) We ask what happens if our knowledge of the sample selection function is imperfect, and potentially biased (Section \ref{sec:results_incompR}). (iii) We investigate how to best account for individual, and possibly misjudged, measurement uncertainties (Section \ref{sec:results_errors}). (iv) We determine which of several stellar sub-populations is best for constraining the potential (Section \ref{sec:results_temperature}). 

One of the strongest assumptions is restricting the dynamical modelling to a certain family of parametrized functions for the gravitational potential and the DF. We investigate how well we can hope to recover the true potential, when our models do not encompass the true DF (Section \ref{sec:results_mixedDFs}) and potential (Section \ref{sec:results_potential}).

For all of the above aspects we show some plausible and illustrative examples on the basis of investigating mock data. The mock data is generated from galaxy models outlined in Sections \ref{sec:coordinates}-\ref{sec:qDF} following the procedure in Appendices \ref{app:mockdata}-\ref{app:selectionfunction} and analysed according to the description of the \RM{} machinery in Sections \ref{sec:data_likelihood}-\ref{sec:fitting}. Section \ref{sec:results} compiles our results on the investigated modelling aspects. In particular, our key results about the systematics introduced by using wrong DF or potential models are presented in the Sections \ref{sec:results_mixedDFs} and \ref{sec:results_potential}. Section \ref{sec:discussionsummary} finally summarizes and discusses our findings.



\section{Dynamical modelling}

In this section we summarize the basic elements of \RM{}, the dynamical modelling machinery presented in this work, which in many respects follows BR13 and makes extensive use of the \texttt{galpy} Python package for galactic dynamics\footnote{\texttt{galpy} is an open-source code that is being developed on \url{http://github.com/jobovy/galpy}. The latest documentation can be found at \url{http://galpy.readthedocs.org/en/latest/}.} \citep{2015ApJS..216...29B}.


\subsection{Coordinate system} \label{sec:coordinates}

Our modelling takes place in the Galactocentric rest-frame with cylindrical coordinates $\vect{x} \equiv (R,\phi,z)$ and corresponding velocity components $\vect{v} \equiv (v_R,v_T,v_z)$. If the stellar phase-space data is given in observed heliocentric coordinates, position $\tilde{\vect{x}} \equiv(\text{RA},\text{Dec},m-M)$ in right ascension RA, declination Dec and distance modulus $(m-M)$, and velocity $\tilde{\vect{v}} \equiv (\mu_\text{RA} \cdot \cos ( \text{Dec}),\mu_\text{Dec},v_\text{los})$ as proper motions and line-of-sight velocity, the data $(\tilde{\vect{x}},\tilde{\vect{v}})$ has to be converted into the Galactocentric rest-frame coordinates $(\vect{x},\vect{v})$ using the Sun's position and velocity. We assume for the Sun
\begin{eqnarray*}
(R_\odot,\phi_\odot,z_\odot) &=&(8~\text{kpc}, 0^\circ, 0~\text{kpc})\\
(v_{R\odot},v_{T\odot},v_{z\odot}) &=& (0,230,0)~\text{km s}^{-1}.
\end{eqnarray*}


\begin{deluxetable*}{llcll}[!htbp]
\tabletypesize{\scriptsize}
\tablecaption{Axisymmetric gravitational potential models used throughout this work. The potential parameters are fixed for the mock data creation at the values given in this table, which we susequently aim to recover with \RM{}. The parameters of \texttt{DHB-Pot} and \texttt{KKS-Pot} were chosen to resemble the \texttt{MW14-Pot} (see Figure \ref{fig:ref_pots}). We use $v_\text{circ}(R_\odot)$ = $230~\text{km s}^{-1}$ as the circular velocity at the Sun for all potentials in this work. \label{tbl:referencepotentials}}
\tablewidth{0pt}
\tablehead{
\colhead{name} & \colhead{potential model} & \multicolumn{2}{c}{parameters $p_\Phi$} & \colhead{action calculation}}
\startdata
\texttt{Iso-Pot} & isochrone potential\tablenotemark{(a)} & $b$& $0.9~\text{kpc}$ & \textbf{\emph{analytic and exact}} \\
 & \citep{1959AnAp...22..126H} & & & (\citealt{2008gady.book.....B}, \S 3.5.2) \\
\tableline
\texttt{KKS-Pot} & 2-component & $\Delta$ & $0.3$ & \textbf{\emph{exact}} \\
 & Kuzmin-Kutuzov- & $\left(\frac{a}{c}\right)_\text{Disk}$ & $20$ & using interpolation\\ 
 & St\"{a}ckel potential\tablenotemark{(b)}: & $\left(\frac{a}{c}\right)_\text{Halo}$ & $1.07$ & on action grid \\
 & disk and halo & $k$ & $0.28$ & \citep{2012MNRAS.426.1324B,2015ApJS..216...29B}\\
 & \citep{1994AA...287...43B} & & & \\
 & & & & \\
 \tableline
\texttt{DHB-Pot} & Disk+Halo+Bulge potential\tablenotemark{(c)}: & $a_\text{disk}$ & $3~\text{kpc}$ & \textbf{\emph{approximate}} \\ 
 & Miyamoto-Nagai disk, & $b_\text{disk}$ & $0.28~\text{kpc}$ (fixed) & using \emph{St\"{a}ckel fudge} \\
 & NFW halo, & $f_\text{halo}$ & $0.35/0.95$ & \citep{2012MNRAS.426.1324B} \\
 & Hernquist bulge & $a_\text{halo}$ & $16~\text{kpc}$ (fixed) & and interpolation on action grid \\
 & (same as \texttt{MW14-Pot}, & $f_\text{bulge}$ & $0.05/1.0$ (fixed) & \\
 & except of bulge) & $a_\text{bulge}$ & $0.6~\text{kpc}$ (fixed) & \\
\tableline
\texttt{MW14-Pot} & MW-like potential\tablenotemark{(d)}: & & & \textbf{\emph{approximate}} \\
 & Miyamoto-Nagai disk, & & & using \emph{St\"{a}ckel fudge}\\
 & NFW halo, & & & \\
 & cut-off power-law bulge & & & \\
 & \citep{2015ApJS..216...29B} & & & 
\enddata
\tablenotetext{(a)}{The free parameter of the spherical \texttt{Iso-Pot} is the isochrone scale length $b$.} 
\tablenotetext{(b)}{The coordinate system of each of the two St\"{a}ckel-potential components of the \texttt{KKS-Pot} is $R^2 / (\tau_{i,p}+\alpha_p) + z^2 / (\tau_{i,p}+\gamma_p)=1$ with $p \in \{\text{Disk},\text{Halo}\}$ and $\tau_{i,p} \in \{\lambda_p,\nu_p\}$. Both components have the same focal distance $\Delta \equiv \sqrt{\gamma_p-\alpha_p}$, to ensure that the superposition itself is a St\"{a}ckel potential. The axis ratio of the coordinate surfaces $\left(a/c\right)_p := \sqrt{\alpha_p /\gamma_p}$ describes the flatness of each component. $k$ is the relative contribution of the disk mass to the total mass.} 
\tablenotetext{(c)}{The parameters of the \texttt{DHB-Pot} are the Miyamoto-Nagai disk scale length $a_\text{disk}$ and height $b_\text{disk}$, the NFW halo scale length $a_\text{halo}$ and its relative contribution to $v_\text{circ}^2(R_\odot)$ with respect to the total disk+halo contribution, $f_\text{halo}$, and the Hernquist bulge scale length $a_\text{bulge}$ and its contribution to the total $v_\text{circ}^2(R_\odot)$, $f_\text{bulge}$. We keep all except $v_\text{circ}(R_\odot), a_\text{disk}$ and $f_\text{halo}$ fixed to their true values in the analysis.} 
\tablenotetext{(d)}{The \texttt{MWPotential2014} by \citet{2015ApJS..216...29B} (see their Table 1) has $v_\text{circ}(R_\odot)=220~\text{km s}^{-1}$. We use however $v_\text{circ}(R_\odot)=230~\text{km s}^{-1}$.} 
\end{deluxetable*}

\begin{figure*}[!htbp]
\includegraphics[width=\textwidth]{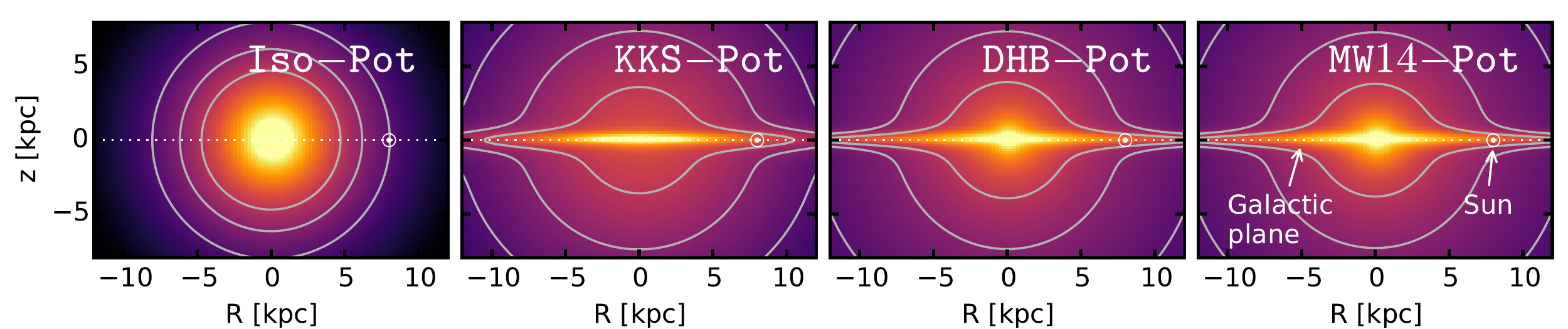}
\caption{Density distribution of the four reference galaxy potentials in Table \ref{tbl:referencepotentials}. These potentials are used throughout this work to create and model mock data with \RM{}.}
\label{fig:ref_pots}
\end{figure*}


\subsection{Actions}

Stellar orbits in (axisymmetric) gravitational potentials are best described and fully specified by the three actions $\vect{J} \equiv (J_R, J_z, J_\phi=L_z)$, defined as
\begin{equation}
J_i \equiv \frac{1}{2\pi} \oint_\text{orbit} p_i \diff x_i, \label{eq:action_general}
\end{equation}
which is evaluated along the orbit with position $\vect{x}(t)$ and momentum $\vect{p}(t)$ in a given potential $\Phi$. Actions have several convenient properties which make them excellent orbit labels and ideal as arguments for orbit DFs: Actions are integrals of motion; actions have an intuitive physical meaning as they quantify the amount of oscillation of the orbit in each coordinate direction; actions---together with a set of angle coordinates $\vect{\theta}$---form canonical conjugate phase-space coordinates, i.e., the Jacobian determinant $\left| \partial(\vect{J},\vect{\theta}) / \partial (\vect{x},\vect{v}) \right|=1$ (with Cartesian $\vect{x}$ and $\vect{v}$). The angles $\vect{\theta}(t)\propto t$ evolve linearly in time and specify the position of the star along the orbit. (For a full introduction to angle-action variables see \citealt{2008gady.book.....B}, \S 3.5.)

Action calculation from a star's phase-space coordinates, $(\vect{x},\vect{v}) \stackrel{\Phi}{\longrightarrow} \vect{J}$, is typically very computationally expensive. Only for some special, separable potentials does Equation \eqref{eq:action_general} simplify significantly. The triaxial St\"{a}ckel potentials \citep{1985MNRAS.216..273D} are the most general potentials, that allow exact action calculations using a single quadrature. Some flattened axisymmetric St\"{a}ckel potentials are quite similar to our Galaxy's potential (\citealt{2008gady.book.....B}, \S 3.5.3; \citealt{1994AA...287...43B,2003MNRAS.340..752F}). The spherical isochrone potential (\citealt{1959AnAp...22..126H,2008gady.book.....B}, \S 3.5.3) is the most general special case for which the action calculation is analytic without any integration. In all other potentials actions have to be numerically estimated; see \citet{2016MNRAS.457.2107S} for a recent review of action estimation methods. According to \citet{2016MNRAS.457.2107S} the best compromise of speed and accuracy for the Galactic disk is the \emph{St\"{a}ckel fudge} by \citet{2012MNRAS.426.1324B} for axisymmetric potentials. In addition we use action interpolation grids \citep{2012MNRAS.426.1324B,2015ApJS..216...29B} to speed up the calculation. The latter is one of the improvements employed by \RM{}, which was not used in BR13. 

\newpage
\subsection{Potential models} \label{sec:potentials}

For the gravitational potential in our modelling we assume a family of parametrized models. We use: A MW-like potential with disk, halo and bulge (\texttt{DHB-Pot}); the spherical isochrone potential (\texttt{Iso-Pot}); and the 2-component Kuzmin-Kutuzov St\"{a}ckel potential (\citealt{1994AA...287...43B}; \texttt{KKS-Pot}), which also displays a disk and halo structure. Table \ref{tbl:referencepotentials} summarizes all reference potentials used in this work together with their free parameters $p_\Phi$. The true circular velocity at the Sun was chosen to be $v_\text{circ}(R_\odot)=230~\text{km s}^{-1}$ for all potential models. The \texttt{Iso-Pot} allows both accurate and particularly fast action calculations; we use it therefore for tests requiring a large number of analyses. The \texttt{KKS-Pot} and \texttt{DHB-Pot} were chosen for their more realistic galaxy shape and because their closed-form expression for $\Phi(R,z)$ makes the computation of forces and densities fast and easy. The \texttt{KKS-Pot} allows also for exact action calculations, while the \texttt{DHB-Pot} has physically more intuitive potential parameters, but requires the \emph{St\"{a}ckel fudge} to estimate actions. The parameter values of \texttt{KKS-Pot} and \texttt{DHB-Pot} in Table \ref{tbl:referencepotentials} were chosen to resemble the MW potential from \citet{2015ApJS..216...29B} (\texttt{MW14-Pot}). The density distribution of all these potentials is illustrated in Figure \ref{fig:ref_pots}. 


\begin{deluxetable*}{lccccc}[!htbp]
\tabletypesize{\scriptsize}
\tablecaption{Reference parameters for the qDF in Equations \eqref{eq:df_general}-\eqref{eq:sigmazRg}, used to create 6D phase-space mock data sets for stellar populations of different kinematic temperature. (The parameter $X$ is explained in Section \ref{sec:qDF}.) \label{tbl:referenceMAPs}}
\tablewidth{0pt}
\tablehead{
\colhead{name} & \multicolumn{5}{c}{qDF parameters $p_\text{DF}$}\\
 & \colhead{$h_R$ [kpc]} & \colhead{$\sigma_{R,0}$ [$\text{km s}^{-1}$]} & \colhead{$\sigma_{z,0}$ [$\text{km s}^{-1}$]} & \colhead{$h_{\sigma,R}$ [kpc]} & \colhead{$h_{\sigma,z}$ [kpc]}}
\startdata
\texttt{hot} & 2 & 55 & 66 & 8 & 7\\
\texttt{cool} & 3.5 & 42 & 32 & 8 & 7\\
\tableline
\texttt{cooler} & $3$ & $27.5$ & $33$ & $8$ & $7$ \\
\texttt{colder} & $2 +X \%$ & $55-X \%$ & $66-X \%$ & $8$ & $7$ \\
\texttt{warmer} & $3.5-X \%$ & $42+X \%$ & $32+X \%$ & $8$ & $7$
\enddata
\end{deluxetable*}

\subsection{Stellar distribution functions} \label{sec:qDF}

A stellar distribution function $\text{DF}(\vect{x},\vect{v})$ can be considered as the probability of a star to be found at $(\vect{x},\vect{v})$. Using instead orbit DFs in terms of $(\vect{J},\vect{\theta})$ has the advantage that the distribution of stars in $\vect{\theta}$ is uniform and the orbit DF reduces effectively to a function of the actions $\vect{J}$ only. As $\left| \partial(\vect{J},\vect{\theta}) / \partial (\vect{x},\vect{v}) \right| = 1$, the function $\text{DF}(\vect{J})$ can still be thought of as a probability in $(\vect{x},\vect{v})$.

The action-based quasi-isothermal distribution function (qDF) by \citet{2010MNRAS.401.2318B} and \citet{2011MNRAS.413.1889B} is a simple DF which we will employ as a specific example throughout this work to describe individual stellar sub-populations. This is motivated by the findings of \citet{2012ApJ...751..131B,2012ApJ...755..115B,2012ApJ...753..148B} and \citet{2013MNRAS.434..652T} on the simple phase-space structure of stellar \MAPs{} and BR13's successful application. The qDF has the form
\begin{eqnarray}
&&\text{qDF}(\vect{J} \mid p_\text{DF}) \nonumber\\
&&= f_{\sigma_R}\left(J_R,L_z \mid p_\text{DF}\right) \times f_{\sigma_z}\left(J_z,L_z \mid p_\text{DF}\right)\label{eq:df_general}\end{eqnarray}
with some free parameters $p_\text{DF}$ and
\begin{eqnarray}
f_{\sigma_R}\left(J_R,L_z \mid p_\text{DF}\right) &=& n \times \frac{\Omega}{\pi\sigma_R^2(R_g) \kappa}\exp\left(-\frac{\kappa J_R}{\sigma_R^2(R_g)} \right) \nonumber\\
&& \times \left[1+\tanh\left(L_z/L_0\right) \right]\\
f_{\sigma_z}\left(J_z,L_z \mid p_\text{DF} \right) &=& \frac{\nu}{2 \pi \sigma_z^2(R_g)} \exp\left( -\frac{\nu J_z}{\sigma_z^2(R_g)} \right)
\end{eqnarray}
\citep{2011MNRAS.413.1889B}. Here $R_g$, $\Omega$, $\kappa$ and $\nu$ are functions of $L_z$ and denote respectively the guiding-center radius, circular, radial/epicycle and vertical frequency of the near-circular orbit with angular momentum $L_z$ in a given potential. The term $\left[1+\tanh\left(L_z/L_0\right) \right]$ suppresses counter-rotation for orbits in the disk with $L_z < L_0$ (with $L_0 \sim 10~\text{km s}^{-1}~ \text{kpc}$).

Following BR13, we choose the functional forms
\begin{eqnarray}
n(R_g \mid p_\text{DF}) &\propto& \exp\left(-\frac{R_g}{h_R} \right)\\
\sigma_R(R_g \mid p_\text{DF}) &=& \sigma_{R,0} \times \exp\left(- \frac{R_g-R_\odot}{h_{\sigma,R}} \right)\label{eq:sigmaRRg}\\
\sigma_z(R_g \mid p_\text{DF}) &=& \sigma_{z,0} \times \exp\left(- \frac{R_g-R_\odot}{h_{\sigma,z}} \right)\label{eq:sigmazRg},
\end{eqnarray}
which indirectly set the stellar number density and radial and vertical velocity dispersion profiles. The qDF has therefore a set of five free parameters $p_\text{DF}$: the density scale length of the tracers $h_R$, the radial and vertical velocity dispersion at the Solar position $R_\odot$, $\sigma_{R,0}$ and $\sigma_{z,0}$, and the scale lengths $h_{\sigma,R}$ and $h_{\sigma,z}$, that describe the radial decrease of the velocity dispersion. \RM{} allows to fit any number of DF parameters simultaneously, while BR13 kept $\{\sigma_{R,0},h_{\sigma,R}\}$ fixed. Throughout this work we make use of a few example stellar populations whose qDF parameters are given in Table \ref{tbl:referenceMAPs}: Most tests use the \texttt{hot} and \texttt{cool} qDFs, which correspond to kinematically hot and cool populations, respectively. The \texttt{warmer} (\texttt{cooler} and \texttt{colder}) qDFs in Table \ref{tbl:referenceMAPs} were chosen to have the same anisotropy $\sigma_{R,0}/\sigma_{z,0}$ as the \texttt{cool} (\texttt{hot}) qDF, with $X$ being a free parameter describing the temperature difference. Hotter populations have shorter tracer scale lengths \citep{2012ApJ...753..148B} and the velocity dispersion scale lengths were fixed according to \citet{2012ApJ...755..115B}.

\begin{figure*}[!htbp]
\centering
\plotone{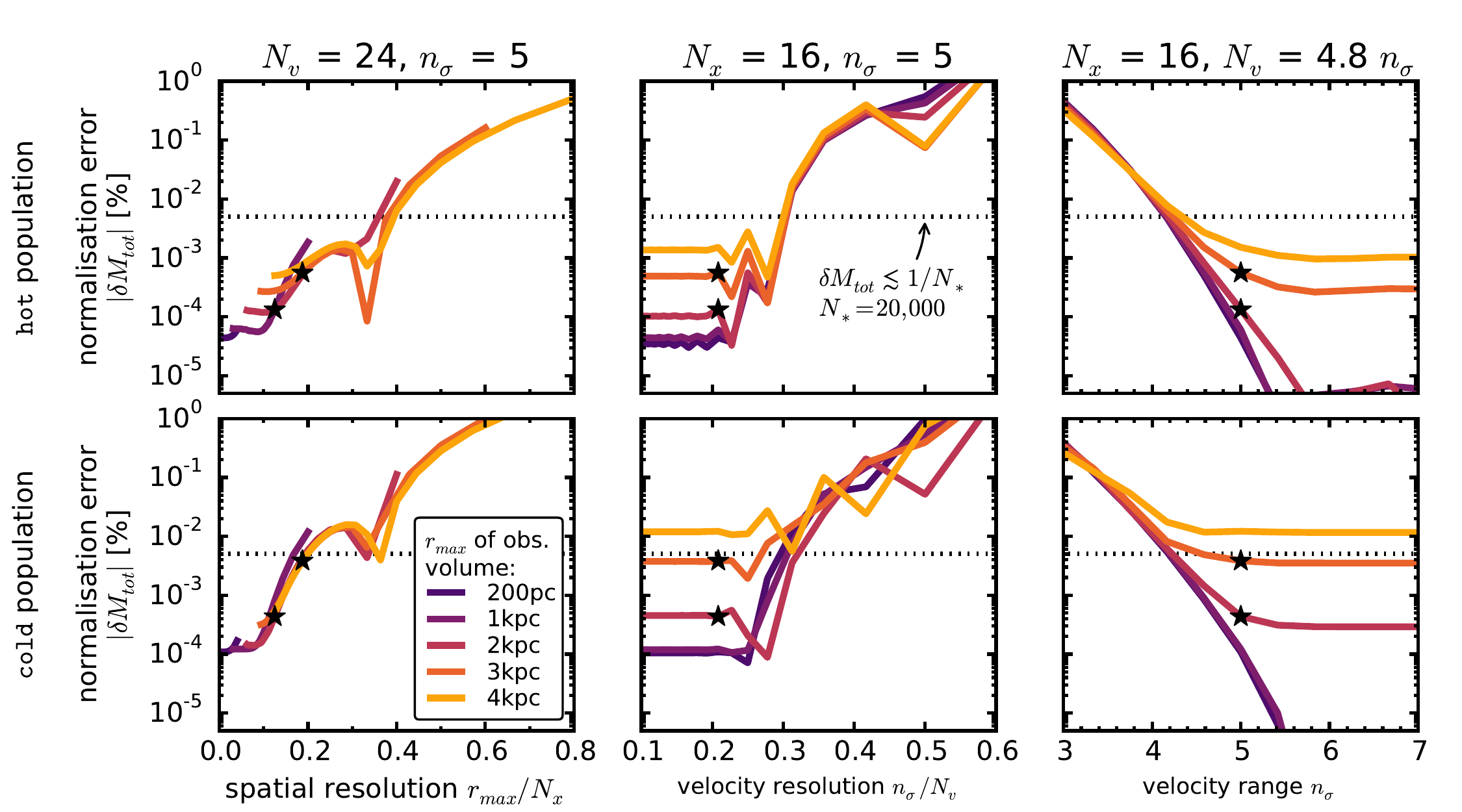}
\caption{Relative error of the likelihood normalisation, $\delta M_\text{tot}$ in Equation \eqref{eq:relerrlikelihood}, depending on the accuracy of the grid-based density calculation in Equation \eqref{eq:tracerdensity} (and surrounding text) in five spherical observation volumes with different radius $r_\text{max}$ and depending on the kinematic temperature of the population in the \texttt{DHB-Pot}. (Test \ref{test:norm_accuracy} in Table \ref{tbl:tests} summarizes the model parameters.) The tracer density in Equation \eqref{eq:tracerdensity} is calculated on $N_x\times N_x$ spatial grid points in $R \in [R_\odot \pm r_\text{max}]$ and $|z| \in [0,r_\text{max}]$. The integration over the velocities is performed with Gauss-Legendre quadratures of order $N_v$ within an integration range of $\pm n_\sigma$ times the dispersion $\sigma_R(R)$ and $\sigma_z(R)$ (and $[0,1.5v_\text{circ}]$ in $v_T$). (We vary $N_x$, $N_v$ and $n_\sigma$ separately and keep the other two fixed at the values indicated above each panel.) We calculate the ``true'' normalisation $M_\text{tot}$ in Equation \eqref{eq:relerrlikelihood} with high accuracy as $M_\text{tot} \equiv M_\text{tot,approx}(N_x=32,N_v=68,n_\sigma=7)$. The black stars indicate the accuracy used in analyses with the \texttt{DHB-Pot}, Tests \ref{test:isoSphFlexIncomp} and \ref{test:MWdhbMix}: It is better than $0.005\%$ (dotted line), which is required for $N_*=20,000$ stars. We find that the spatial resolution of the grid is important and depends on the kinematic temperature of the population, as cooler populations have a steeper density gradient in $z$-direction, which has to be sampled sufficiently.}
\label{fig:norm_accuracy}
\end{figure*}

One indispensable step in our dynamical modelling technique (Section \ref{sec:data_likelihood}-\ref{sec:likelihood_normalisation}), as well as in creating mock data (Appendix \ref{app:mockdata}), is to calculate the (axisymmetric) spatial tracer density $\rho_\text{DF}(\vect{x} \mid p_{\Phi},p_\text{DF})$ for a given DF and potential. Analogously to BR13, 
\begin{eqnarray}
&&\rho_\text{DF}(R,|z| \mid p_{\Phi},p_\text{DF}) \nonumber\\
&&= \int_{-\infty}^{\infty} \text{DF}(\vect{J}[R,z,\vect{v} \mid p_{\Phi}] \mid p_\text{DF}) \Diff3 v \nonumber\\
&&\approx \int_{-n_\sigma \sigma_R(R \mid p_\text{DF})}^{n_\sigma \sigma_R(R \mid p_\text{DF})} \int_{-n_\sigma\sigma_z(R \mid p_\text{DF})}^{n_\sigma \sigma_z(R \mid p_\text{DF})} \int_{0}^{1.5 v_\text{circ}(R_\odot)} \nonumber\\
& & \hspace{1cm} \text{DF}(\vect{J}[R,z,\vect{v} \mid p_{\Phi}] \mid p_\text{DF}) \diff v_T \diff v_z \diff v_R, \label{eq:tracerdensity}
\end{eqnarray}
where $\sigma_R(R \mid p_\text{DF})$ and $\sigma_z(R \mid p_\text{DF})$ are given by Equations \eqref{eq:sigmaRRg} and \eqref{eq:sigmazRg}.\footnote{The integration ranges over the velocities are motivated by Figure \ref{fig:kks2WedgeEx_xv} and $n_\sigma$ should be chosen as $n_\sigma \sim 5$ (see Figure \ref{fig:norm_accuracy}). The integration range $[0,1.5 v_\text{circ}(R_\odot)]$ over $v_T$ is in general sufficient, only for observation volumes with larger mean stellar $v_T$ this upper limit needs to be increased.} Each integral is evaluated using a $N_v$-th order Gauss-Legendre quadrature. For a given $p_\Phi$ and $p_\text{DF}$ we explicitly calculate the density on $N_x \times N_x$ regular grid points in the $(R,z)$ plane and interpolate $\ln \rho_\text{DF}$ in between using bivariate spline interpolation. The grid is chosen to cover the extent of the observations (for $|z|\geq0$, because the model is symmetric in $z$ by construction). The total number of actions to be calculated to set up the density interpolation grid is $N_x^2 \times N_v^3$, which is one of the factors limiting the computation speed. To complement the work by BR13, we will specifically work out in Section \ref{sec:likelihood_normalisation} and Figure \ref{fig:norm_accuracy} how large $N_x$, $N_v$ and $n_\sigma$ have to be chosen to get the density with a sufficiently high numerical accuracy. 


\subsection{Data likelihood} \label{sec:data_likelihood}

As data $D$ we consider here the positions and velocities of a sub-population of stars within a given survey selection function $\text{SF}(\vect{x})$,
\begin{eqnarray*}
D \equiv \{ \vect{x}_i,\vect{v}_i \mid && \text{(star $i$ in given sub-population)}\nonumber\\
&\wedge& (\text{SF}(\vect{x}_i) > 0) \}.
\end{eqnarray*}

For simplicity we assume in most tests of this study contiguous, spherical SFs centred on the Sun, which are functions of $\vect{x}$ only and which we motivate in Appendix \ref{app:selectionfunction}. The maximum radius of this spherical observed volume is denoted by $r_\text{max}$.

We fit a model potential and DF (here: the qDF) which are specified by a number of fixed and free model parameters,
\begin{eqnarray*}
\pmodel \equiv \{ p_\text{DF} , p_\Phi \}.
\end{eqnarray*}
The orbit of the $i$-th star in a potential with $p_\Phi$ is labelled by the actions $\vect{J}_i \equiv \vect{J}[\vect{x}_i,\vect{v}_i\mid p_{\Phi}]$ and the DF evaluated for the $i$-th star is then $\text{DF}(\vect{J}_i \mid \pmodel) \equiv \text{DF}(\vect{J}[\vect{x}_i,\vect{v}_i\mid p_{\Phi}] \mid p_\text{DF})$.

The likelihood of the data given the model is, following BR13 and \citet{2013MNRAS.433.1411M},
\begin{eqnarray}
&&\mathscr{L}(D \mid \pmodel) \nonumber\\
&&\equiv \prod_i^{N_*} p(\vect{x}_i,\vect{v}_i \mid \pmodel) \nonumber\\
&&= \prod_i^{N_*} \frac{\text{DF}(\vect{J}_i \mid \pmodel) \cdot \text{SF}(\vect{x}_i)}{\int \text{DF}(\vect{J} \mid \pmodel) \cdot \text{SF}(\vect{x}) \Diff 3 x \Diff 3 v}\nonumber\\
&&\propto \prod_i^{N_*} \frac{\text{DF}(\vect{J}_i \mid \pmodel)}{\int \rho_\text{DF}(R,|z| \mid \pmodel) \cdot \text{SF}(\vect{x}) \Diff 3 x}, \label{eq:prob}
\end{eqnarray}
where $N_*$ is the number of stars in $D$, and in the last step we used Equation \eqref{eq:tracerdensity}.\footnote{Because $\left| \partial(\vect{J},\vect{\theta}) / \partial (\vect{x},\vect{v}) \right| = 1$, the integration over phase-space in the normalisation term can be performed either over $(\vect{J},\vect{\theta})$ or Cartesian $(\vect{x},\vect{v})$.} $\prod_i^{N_*}\text{SF}(\vect{x}_i)$ is independent of \pmodel{}, so we treat it as unimportant proportionality factor. We find the best fitting \pmodel{} by maximizing the posterior probability distribution $\pdf{}(\pmodel \mid D)$, which is, according to Bayes' theorem
\begin{equation*}
\pdf{}(\pmodel \mid D) \propto \mathscr{L}(D\mid \pmodel) \cdot p(\pmodel),
\end{equation*}
where $p(\pmodel)$ is some prior probability distribution on the model parameters. We assume flat priors in both $p_\Phi$ and
\begin{eqnarray}
p_\text{DF} := \{ \ln h_R, \ln \sigma_{R,0}, \ln \sigma_{z,0}, \ln h_{\sigma,R}, \ln h_{\sigma,z} \} \label{eq:p_DF}
\end{eqnarray}
(see Section \ref{sec:qDF}) throughout this work. Then \pdf{} and likelihood are proportional to each other and differ only in units.

In this case, where we use uninformative priors, a maximum-likelihood estimation procedure (e.g., via the expectation-maximization (EM) algorithm and parameter uncertainty estimates from the Fisher information matrix) would lead to the same result as the Bayesian inferential procedure described in this work (see Section \ref{sec:fitting}). We expect however that in due course increasingly informative priors will become available (like, e.g., rotation curve measurements from maser sources by \citet{2009ApJ...700..137R}) and Bayesian inference is therefore the preferred framework.


\subsection{Likelihood normalisation} \label{sec:likelihood_normalisation}

The normalisation in Equation \eqref{eq:prob} is a measure for the total number of tracers inside the survey volume,
\begin{equation}
M_\text{tot} \equiv \int \rho_\text{DF}(R,|z| \mid \pmodel) \cdot \text{SF}(\vect{x}) \Diff 3 x.\label{eq:normalisation}
\end{equation}
In the case of an axisymmetric Galaxy model and $\text{SF}(\vect{x})=1$ within the observation volume (as in most tests in this work), the normalisation is essentially a two-dimensional integral in the $(R,z)$ plane over $\rho_{DF}$ with finite integration limits. We evaluate the integrals using Gauss-Legendre quadratures of order 40. The integral over the azimuthal direction can be solved analytically. 

It turns out that a sufficiently accurate evaluation of the likelihood is computationally expensive, even for only one set of model parameters. This expense is dominated by the number of action calculations required, which in turn depends on $N_*$ and the numerical accuracy of the tracer density interpolation grid with $N_x^2 \times N_v^3$ grid points in Equation \eqref{eq:tracerdensity} needed for the likelihood normalisation in Equation \eqref{eq:normalisation}. The accuracy of the normalisation has to be chosen high enough, such that the resulting numerical error 
\begin{equation}
\delta M_\text{tot} \equiv \frac{M_\text{tot,approx}(N_x,N_v,n_\sigma) - M_\text{tot} }{M_\text{tot}}\label{eq:relerrlikelihood}
\end{equation}
does not dominate the numerically calculated log-likelihood, i.e.,
\begin{eqnarray}
& & \ln \mathscr{L}_\text{approx}(D \mid \pmodel) \nonumber\\
&& = \sum_i^{N_*} \ln \text{DF}(\vect{J_i} \mid \pmodel) - N_* \ln(M_\text{tot})\nonumber\\
&& - N_* \ln (1 + \delta M_\text{tot}),\label{eq:loglikelihood_relerr}
\end{eqnarray}
with
\begin{eqnarray}
\ln (1 + \delta M_\text{tot}) \leq \frac{1}{N_{*}},\label{eq:accuracycondition}
\end{eqnarray}
and therefore $\delta M_\text{tot} \lesssim 1/N_*$. Otherwise numerical inaccuracies could lead to systematic biases in the potential and DF recovery. For data sets as large as $N_* = 20,000$ stars, which in the age of Gaia could very well be the case, one needs a numerical accuracy of 0.005\% in the normalisation. We made sure that this is satisfied for all analyses in this work. Figure \ref{fig:norm_accuracy} demonstrates how the numerical accuracy for analyses with the \texttt{DHB-Pot} depends on the spatial and velocity resolution of the grid and that the accuracy we use, $N_x=16$, $N_v=24$ and $n_\sigma=5$, is sufficient.\footnote{The accuracy used in this work's analyses is slightly higher than in BR13, where $N_*$ was only a few $\sim 100$.} It has to be noted however, that the optimal values for $N_x$, $N_v$ and $n_\sigma$ depend not only on $N_*$, but also on the kinematic temperature of the population (and to a certain degree even on the choice of potential\footnote{In Figure \ref{fig:MW14vsKKS2New_violins} we will show a comparison for the qDF parameters of two very similar mock data distributions in two different potentials, the \texttt{MW14-Pot} and a best fit potential of the parametric form of the \texttt{KKS-Pot}. As some of the qDF parameters in both potentials are very different, and even more different from the actual physical scale lengths and velocity dispersions, an optimal $n_\sigma$ has to be estimated first for a given potential model before running the \RM{} analysis.}) and it has to be checked on a case-by-case basis what the optimal accuracy is. 

\citet{2013MNRAS.433.1411M}, who use a similar modelling approach and likelihood normalisation, argued that the required accuracy for the normalisation scales as $\log_{10} \left(1+\delta M_\text{tot} \right) \leq 1 / N_* \Rightarrow \delta M_\text{tot} \lesssim 2.3/N_*$, which is satisfied for our tests as well. They evaluate the integrals in the normalisation via Monte-Carlo integration with $\sim 10^9$ sample points in action space. Our approach uses a tracer density interpolation grid for which the resolution needs to be optimized by hand, but it has the advantage that it then only requires the calculation of $N_x^2 \times N_v^3 \sim 4\cdot 10^{6} - 10^7$ actions per normalisation.

\begin{figure}[!htbp]
\centering
\includegraphics[width=\columnwidth]{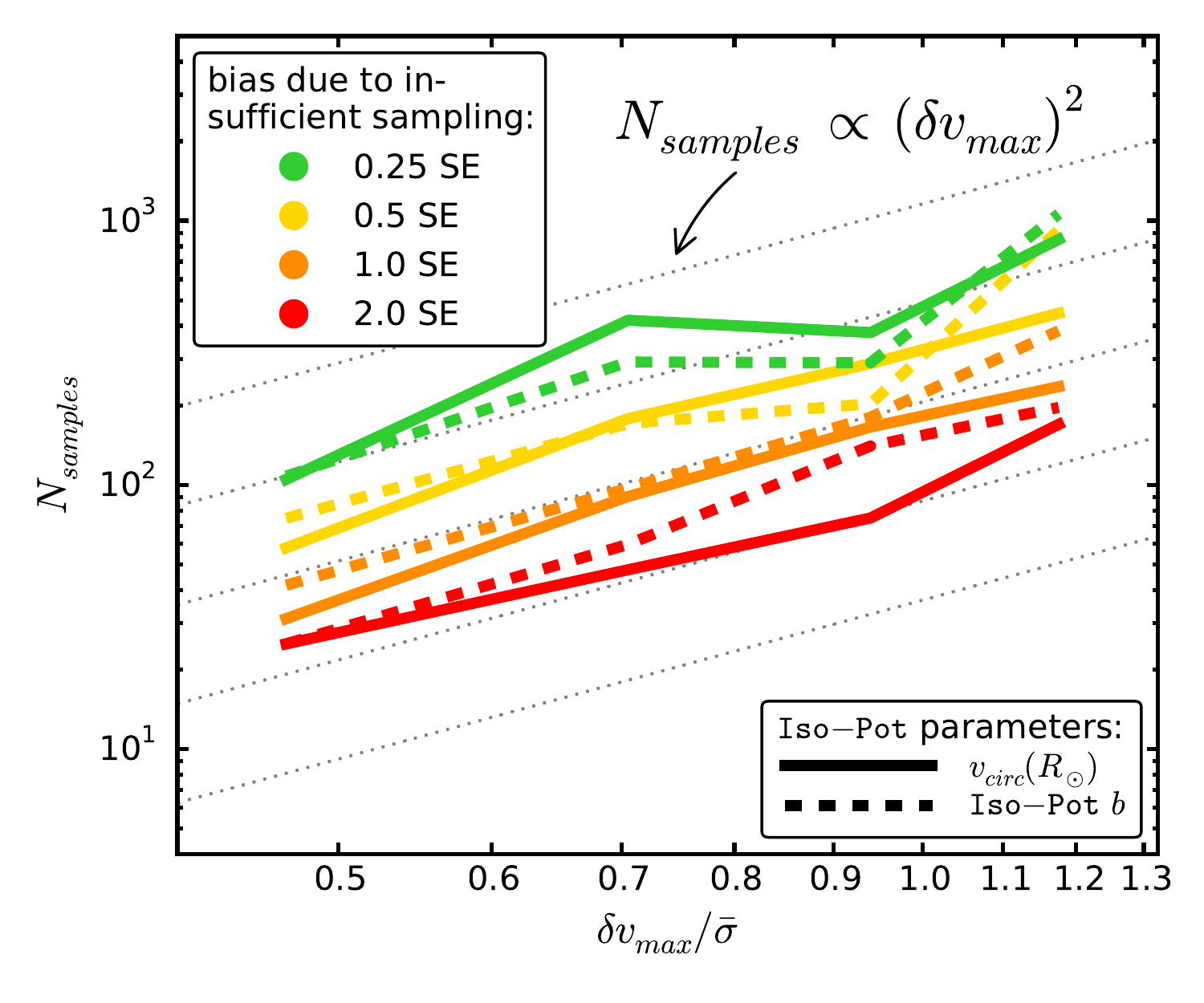}
\caption{Number of MC samples $N_\text{samples}$ needed for the numerical convolution of the model probability with the measurement uncertainties in Equation \eqref{eq:errorconv}, given the maximum velocity uncertainty $\delta v_\text{max}$ within the stellar sample with respect to the sample's kinematic temperature $\bar{\sigma}$. Insufficient sampling introduces systematic biases in the parameter recovery; the size of the bias (in units of the standard error (SE) on the parameter estimate) is indicated in the legend. The relation found here, $N_\text{samples} \propto \delta v_\text{max}^2$, was distilled from analyses (with different $N_\text{samples}$) of mock data sets with different proper motion uncertainties $\delta \mu \in [2,5]~\text{mas yr}^{-1}$ (see Test \ref{test:isoSphFlexErrConv_MC_vs_error} in Table \ref{tbl:tests}). As the reference for the converged convolution integral, we used $N_\text{samples} = 800$ and $1200$ for $\delta \mu \leq 3~\text{mas yr}^{-1}$ and $\delta \mu > 3~\text{mas yr}^{-1}$, respectively (see also left panels in Figure \ref{fig:isoSphFlexErrConv_bias_vs_SE}). We plot $\delta v_\text{max}$ in units of the sample temperature, which we quantify by $\bar{\sigma} \equiv (\sigma_{R,0} + \sigma_{z,0})/2$ (see Table \ref{tbl:referenceMAPs} for the \texttt{hot} qDF).}
\label{fig:isoSphFlexErrConv_MC_vs_error}
\end{figure}

\subsection{Measurement uncertainties}

Measurement uncertainties of the data have to be incorporated in the likelihood. We assume Gaussian uncertainties in the observable space $\vect{y} \equiv (\tilde{\vect{x}},\tilde{\vect{v}})=(\text{RA},\text{Dec},(m-M),\mu_\text{RA} \cdot \cos (\text{Dec}),\mu_\text{Dec},v_\text{los})$, i.e., the $i$-th star's observed $\vect{y}_i$ is drawn from the normal distribution $N[{\vect{y}_i}',\delta \vect{y}_i] \equiv \prod_i^6 N[{y_{i,k}}',\delta y_{i,k}] = \prod_i^6 \exp \{-(y_{k}-{y_{i,k}}')^2/ (2 \delta y_{i,k}^2) \} / \sqrt{2 \pi \delta y_{i,k}^2}$, with ${\vect{y}_i}'$ being the star's true phase-space position, $\delta \vect{y}_i$ its uncertainty, and $y_k$ the $k$-th coordinate component of $\vect{y}$. Stars follow the $\text{DF}(\vect{J}[\vect{y}' \mid p_\Phi] \mid p_\text{DF})$ ($\equiv \text{DF}(\vect{y}') \equiv$ for short) convolved with the measurement uncertainties $N[0,\delta \vect{y}_i]$. The selection function SF$(\vect{y})$ acts on the space of (uncertainty affected) observables. Then the probability of one star becomes
\begin{eqnarray}
&&\tilde{p}(\vect{y}_i \mid p_\Phi,p_\text{DF},\delta \vect{y}_i)\nonumber\\
& \equiv& \frac{\text{SF}(\vect{y}_i) \cdot \int \text{DF}(\vect{y}') \cdot N[\vect{y}_i,\delta \vect{y}_i] \Diff{6} y'}{\int \left( \text{DF}(\vect{y}') \cdot \int \text{SF}(\vect{y}) \cdot N[\vect{y}',\delta \vect{y}_i] \Diff{6} y \right) \Diff{6}y'}.
\end{eqnarray}
In the case of uncertainties in distance and/or $(\text{RA},\text{Dec})$ the evaluation of this is computational very expensive---especially if the stars have heteroscedastic $\delta \vect{y}_i$, which is the case for realistic data sets, and the normalisation needs to be calculated for each star separately. In practice we compute the convolution using Monte Carlo (MC) integration with $N_\text{samples}$ samples,
\begin{eqnarray}
&&\tilde{p}_\text{approx}(\vect{y}_i \mid p_\Phi,p_\text{DF},\delta \vect{y}_i) \nonumber\\
&&\approx \frac{ \text{SF}(\tilde{\vect{x}}_i)}{M_\text{tot}} \cdot \frac{1}{N_\text{samples}} \sum_n^{N_\text{samples}} \text{DF}(\tilde{\vect{x}}_i,\vect{v}[\vect{y}'_{i,n}]) \label{eq:errorconv}
\end{eqnarray}
with
\begin{eqnarray}
\vect{y}'_{i,n} \sim N[\vect{y}_i,\delta \vect{y}_i].\nonumber
\end{eqnarray}

In addition, this approximation assumes that the star's position $\tilde{\vect{x}}_i$ is perfectly measured. As the SF is also velocity independent, this simplifies the normalisation drastically to $M_\text{tot}$ in Equation \eqref{eq:normalisation}. Measurement uncertainties in $\mathrm{RA}$ and $\mathrm{Dec}$ are often negligible anyway. The uncertainties in the Galactocentric velocities $\vect{v}_i = (v_{R,i},v_{T,i},v_{z,i})$ depend not only on $\delta \vect{\mu}$ and $\delta v_\text{los}$ but also on the distance and its uncertainty, which we do \emph{not} neglect when drawing MC samples $\vect{y}'_{i,n}$ from the full uncertainty distribution $N[\vect{y}_i,\delta \vect{y}_i]$. 

An analogous but one-dimensional treatment of measurement uncertainties in only $v_z$ was already applied by BR13. Similar approaches ignoring measurement uncertainties in the likelihood normalisation and using MC sampling of the error ellipses were also used by \citet{2013MNRAS.433.1411M} and \citet{2016MNRAS.tmp..817D}. In Section \ref{sec:results_errors}, Figure \ref{fig:isoSphFlexErrConv_bias_vs_SE} (Test \ref{test:isoSphFlexErrConv_bias_vs_SE} in Table \ref{tbl:tests}), we will investigate the breakdown of our approximation for non-negligible distance uncertainties.

Figure \ref{fig:isoSphFlexErrConv_MC_vs_error} demonstrates that in the absence of position uncertainties the $N_\text{samples}$ needed for the convolution integral to converge depends as
\begin{equation*}
N_\text{samples} \propto \left( \delta v \right)^2
\end{equation*}
on the uncertainties in the (1D) velocities. 
Figure \ref{fig:isoSphFlexErrConv_MC_vs_error} is based on analyses of mock data sets with different proper motion uncertainties $\delta \mu$ (see Test \ref{test:isoSphFlexErrConv_MC_vs_error} in Table \ref{tbl:tests} for all parameters). The proper motion uncertainty $\delta \mu$ translates to heteroscedastic velocity uncertainties according to 
\begin{equation*}
\delta v [\text{km s}^{-1}] \equiv 4.74047 \cdot r[\text{kpc}] \cdot \delta \mu [\text{mas yr}^{-1}],
\end{equation*}
with $r$ being the distance star---Sun. Stars with larger $\delta v$ require more $N_\text{samples}$ for the integral over their measurement uncertainties to converge; Figure \ref{fig:isoSphFlexErrConv_MC_vs_error} therefore shows how the $N_\text{samples}$---needed for the \pdf{} of the \emph{whole} data set to be converged---depends on the \emph{largest} velocity error $\delta v_\text{max} \equiv \delta v(r_\text{max})$ within the data set.

These mock data sets contained each $N_{*}=10,000$ stars. We found that for $N_{*}=5,000$ the required $N_\text{samples}$ to reach a given accuracy becomes smaller for $v_\text{circ}(R_\odot)$, but remains similar for $b$. The former is consistent with our expectation that we need higher accuracy and therefore more $N_\text{samples}$ for larger data sets. The latter seems to be a special property of the \texttt{Iso-Pot} (see also the discussion in Section \ref{sec:results_incompR}).


\begin{figure*}[!htbp]
\centering
\includegraphics[width=0.7\textwidth]{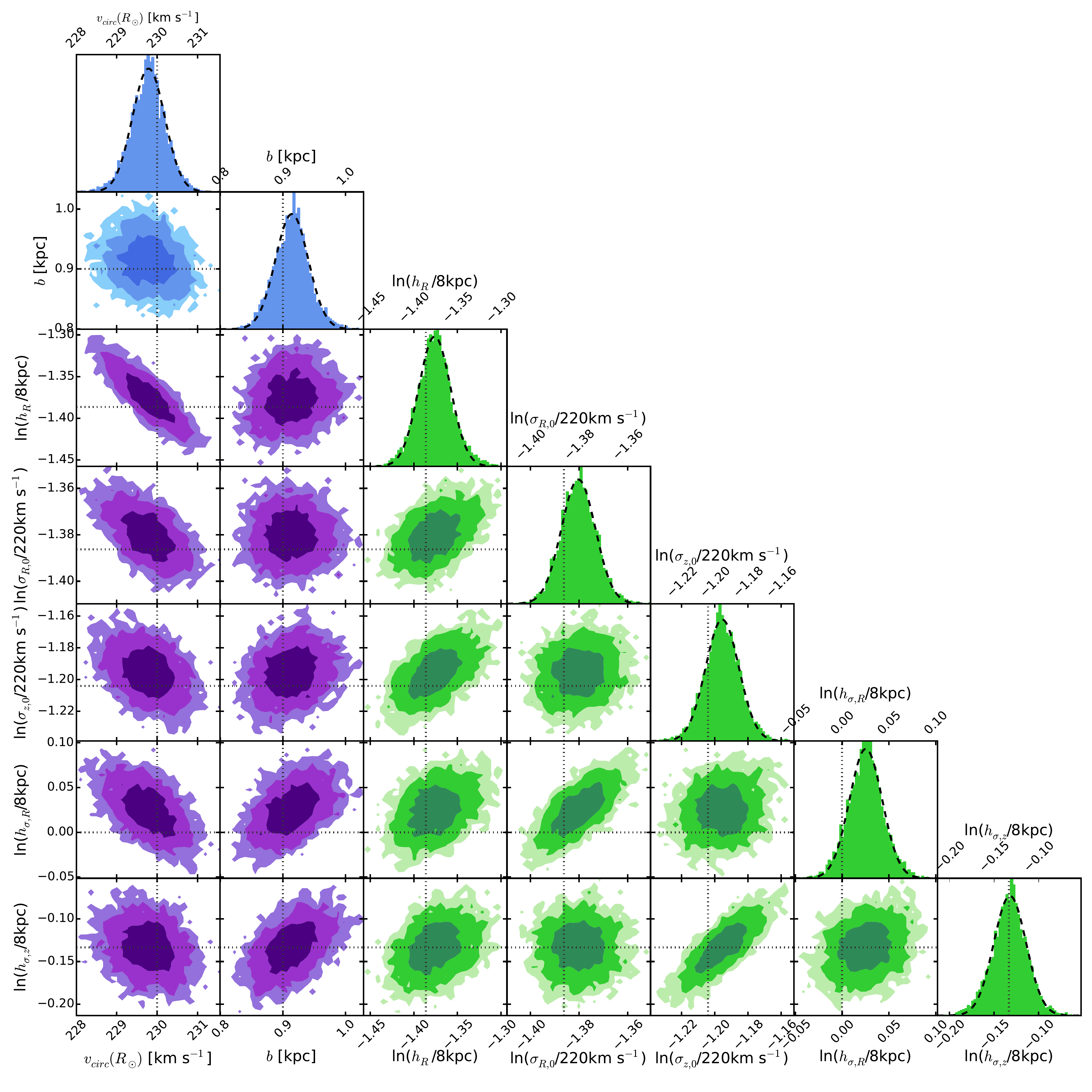}
\caption{The \pdf{} in the parameter space $\pmodel{} = \{p_\Phi,p_\text{DF}\}$ for one example mock data set (see Test \ref{test:isoSphFlex} in Table \ref{tbl:tests}). Blue indicates the \pdf{} for the potential parameters $p_\Phi$, green the qDF parameters $p_\text{DF}$. The true parameters are marked by dotted lines. The dark, medium and bright contours in the 2D distributions represent $1\sigma$, $2\sigma$ and $3\sigma$ confidence regions, respectively. The parameters are weakly to moderately covariant, but their level of covariance depends on the actual choice of the mock data's \pmodel{}. The \pdf{} here was sampled using MCMC. The dashed lines in the 1D distributions are Gaussian fits to the histogram of MCMC samples. This demonstrates that for such a large number of stars, the \pdf{} approaches the shape of a multi-variate Gaussian that also projects into Gaussians when considering the marginalized \pdf{} for all the individual \pmodel{}, as expected for a maximum likelihood estimator.}
\label{fig:isoSphFlex_triangleplot}
\end{figure*}

\subsection{Fitting procedure} \label{sec:fitting}

To search the $(p_\Phi,p_\text{DF})$ parameter space for the maximum of the \pdf{} in Equation \eqref{eq:prob}, we go beyond the single fixed grid search by BR13 and employ an efficient two-step procedure: Nested-grid search and Monte-Carlo Markov Chain (MCMC).

The first step employs a nested-grid search to find the approximate peak and width of the \pdf{} in the high-dimensional \pmodel{} space with a low number of likelihood evaluations:

\begin{itemize}
\item \emph{Initialization.} For $N_p$ free model parameters \pmodel{} we start with a sufficiently large grid with $3^{N_p}$ regular points.

\item \emph{Evaluation.} We evaluate the \pdf{} at each grid-point similar to BR13 (their Figure 9): An outer loop iterates over the potential parameters $p_\Phi$ and pre-calculates all $N_* \times N_\text{samples} + N_x^2 \times N_v^3$ actions required for the likelihood calculation (see Equations \eqref{eq:tracerdensity}, \eqref{eq:prob} and \eqref{eq:errorconv}). Then an inner loop evaluates Equation \eqref{eq:prob} (or \eqref{eq:errorconv}) for all DF parameters $p_\text{DF}$ in the given potential.

\item \emph{Iteration.} For each of the model parameters \pmodel{} we marginalize the \pdf{}. A Gaussian is fitted to the marginalized \pdf{} and the peak $\pm ~ 4\sigma$ become the boundaries of the next grid with $3^{N_p}$ grid points. The grid might be still too coarse or badly positioned to fit Gaussians. In that case we either zoom into the grid point with the highest probability or shift the current range to find new approximate grid boundaries. We proceed with iteratively evaluating the \pdf{} on finer and finer grids, until we have found a reliable $4\sigma$ fit range in each of the \pmodel{} dimensions. The central grid point is then very close to the best fit \pmodel{}, and the grid range is of the order of the \pdf{} width.

\item \emph{The fiducial qDF.} To save time by pre-calculating actions, they have to be independent of the choice of $p_\text{DF}$. However, the normalisation in Equation \eqref{eq:normalisation} requires actions on a $N_x^2 \times N_v^3$ grid and the grid ranges in velocity space \emph{do} depend on the current $p_\text{DF}$ (see Equation \eqref{eq:tracerdensity}). To relax this, we follow BR13 and use a fixed set of qDF parameters (the \emph{fiducial qDF}) to set the velocity grid boundaries in Equation \eqref{eq:tracerdensity} globally for a given $p_\Phi$. Choosing a fiducial qDF that is very different from the true DF can however lead to large biases in the \pmodel{} recovery. BR13 did not account for that. \RM{} avoids this as follows: To get successively closer to the optimal fiducial qDF---with the (yet unknown) best fit $p_\text{DF}$---we use in each iteration step of the nested-grid search the central grid point of the current \pmodel{} grid as the fiducial qDF's $p_\text{DF}$. As the nested-grid search approaches the best fit values, the fiducial qDF approaches its optimum as well. 

\item \emph{Computational expense.} Overall the computation speed of this nested-grid approach is dominated (in descending order of importance) by a) the complexity of potential and action calculation, b) the $N_* \times N_\text{samples} + N_x^2 \times N_v^3$ actions required to be calculated per $p_\Phi$, c) the number of potential parameters and d) the number of DF parameters.
\end{itemize}

The second step samples the shape of the \pdf{} using MCMC. Formally, calculating the \pdf{} on a fine grid like BR13 (e.g., with $K=11$ grid points in each dimension) would provide the same information. However the number of expensive \pdf{} evaluations scales as $K^{N_p}$. For a high-dimensional \pmodel{} ($N_p>4$), a MCMC approach might sample the \pdf{} much faster: We use \emph{emcee} by \citet{2013PASP..125..306F} and release the walkers very close to the best fit \pmodel{} found by the nested-grid search, which assures fast convergence in much less than $K^{N_p}$ \pdf{} evaluations. We also use the best fit \pmodel{} of the grid-search as fiducial qDF for the whole MCMC. In doing so, the normalisation varies smoothly with different $\pmodel{}$ and is less sensitive to the accuracy in Equation \eqref{eq:tracerdensity}.


\section{Results} \label{sec:results}

We are now in a position to examine the limitations of action-based modelling posed in the introduction using our \RM{} machinery. We explore: (i) whether the parameter estimates are unbiased, (ii) the role of the survey volume, (iii) imperfect selection functions, (iv) measurement uncertainties, and what happens if the true (v) DF or (vi) potential are not included in the space of models. 

We will rely on mock data as input to explore the limitations of the modelling. The mock data is generated directly from the fiducial potential and DF models introduced in Sections \ref{sec:potentials} and \ref{sec:qDF}, following the procedure described in Appendix \ref{app:mockdata}. With the exception of the test suite on measurement uncertainties in Section \ref{sec:results_errors}, we assume that phase-space uncertainties are negligible. All tests are also summarized in Table \ref{tbl:tests}. 

\begin{figure}[!htbp]
\centering
\includegraphics[width=\columnwidth]{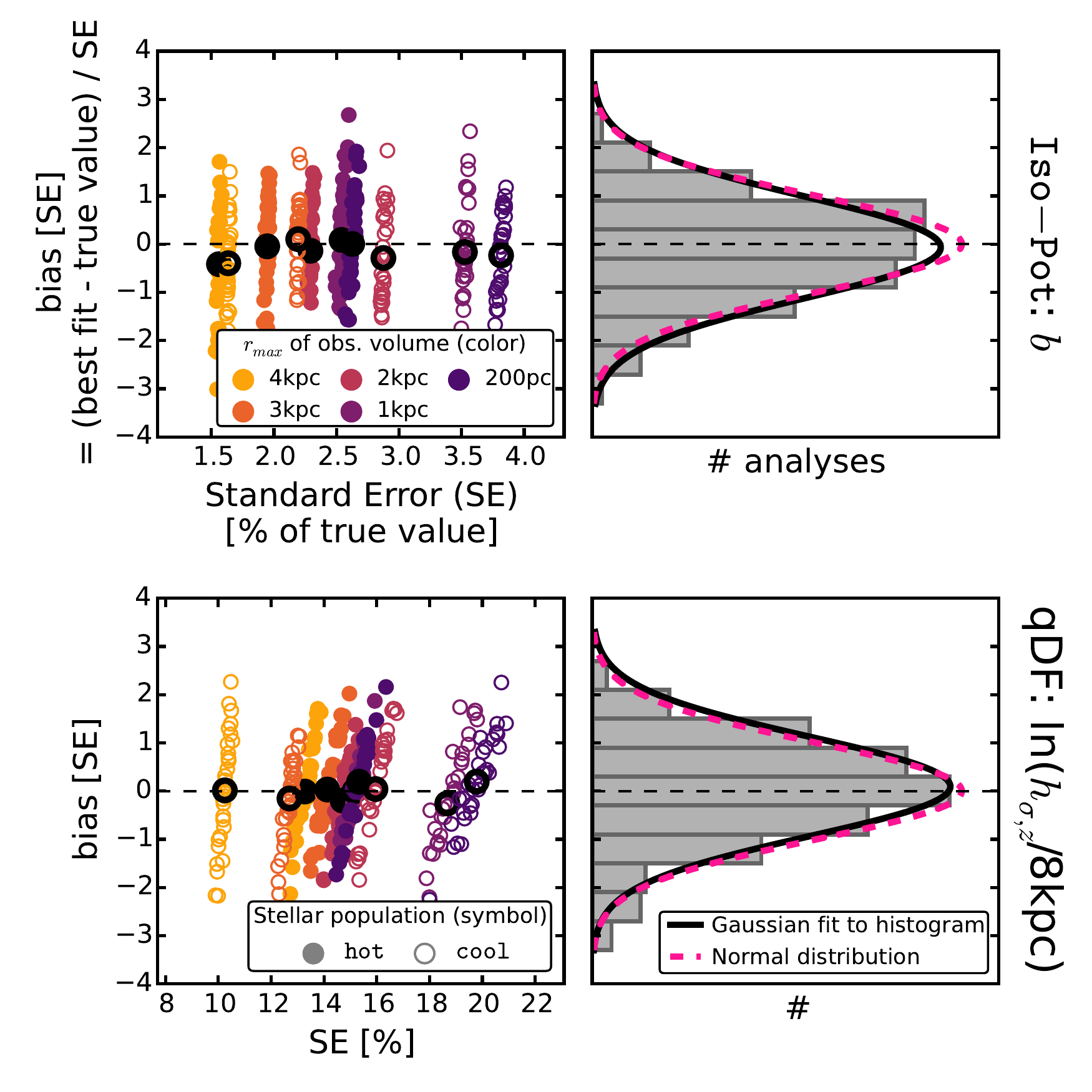}
\caption{Lack of bias in the parameter estimates. Maximum likelihood estimators converge to the true parameter values for large numbers of data points and have a Gaussian spread---if the model assumptions are fulfilled. To test that these conditions are satisfied for \RM{}, we create 320 mock data sets, which come from two different stellar populations and five spherical observation volumes (see legends). (All model parameters are summarized in Table \ref{tbl:tests} as Test \ref{test:isoSph_CLT}.) Bias and relative standard error (SE) are derived from the marginalized \pdf{} for two model parameters (isochrone scale length $b$ in the first row and qDF parameter $h_{\sigma,z}$ in the second row). The second column displays a histogram of the 320 bias offsets. As it closely follows a normal distribution, our modelling method is therefore well-behaved and unbiased. The black dots denote the \pdf{} expectation value for the 32 analyses belonging to the same $\pmodel{}$.}
\label{fig:isoSph_CLT}
\end{figure}

We do not explore the breakdown of the assumption that the system is axisymmetric and in steady state nor the impact of resonances, which is not possible in the current setup using mock data drawn from axisymmetric galaxy models. We plan however to investigate this in a future paper, where we will apply \RM{} to N-body simulations of disk galaxies.


\subsection{Model parameter estimates in the limit of large data sets} \label{sec:largedata}

The \MAP{}s in BR13 contained between 100 and 800 objects, which implied broad \pdf{}s for the model parameters $\pmodel{}$. Several consequences arise in the limit of much larger samples, say $N_{*} = 20,000$: (i) As outlined in Section \ref{sec:likelihood_normalisation} and investigated in Figure \ref{fig:norm_accuracy} (Test \ref{test:norm_accuracy} in Table \ref{tbl:tests}), higher numerical accuracy is needed due to the likelihood normalisation requirement $\delta M_\text{tot} \lesssim 1/N_{*}$ (see Equation \eqref{eq:accuracycondition}), which drives the computing time. (ii) The \pdf{}s of the \pmodel{} become Gaussian, with a \pdf{} width (i.e., the standard error (SE) on the parameter estimate) that scales as $1/\sqrt{N_{*}}$. The former is demonstrated in Figure \ref{fig:isoSphFlex_triangleplot} (Test \ref{test:isoSphFlex} in Table \ref{tbl:tests}) and we also verified that the latter is true. (iii) Any bias in the \pdf{} expectation value has to be considerably less than the SE. Figure \ref{fig:isoSph_CLT} (Test \ref{test:isoSph_CLT} in Table \ref{tbl:tests}) illustrates that \RM{} behaves like an unbiased maximum likelihood estimator: The average parameter estimates from many mock data sets are very close to the input \pmodel{}, and the distribution of the actual parameter estimates are Gaussian around it.

\begin{figure}[!htbp]
\centering
\includegraphics[width=\columnwidth]{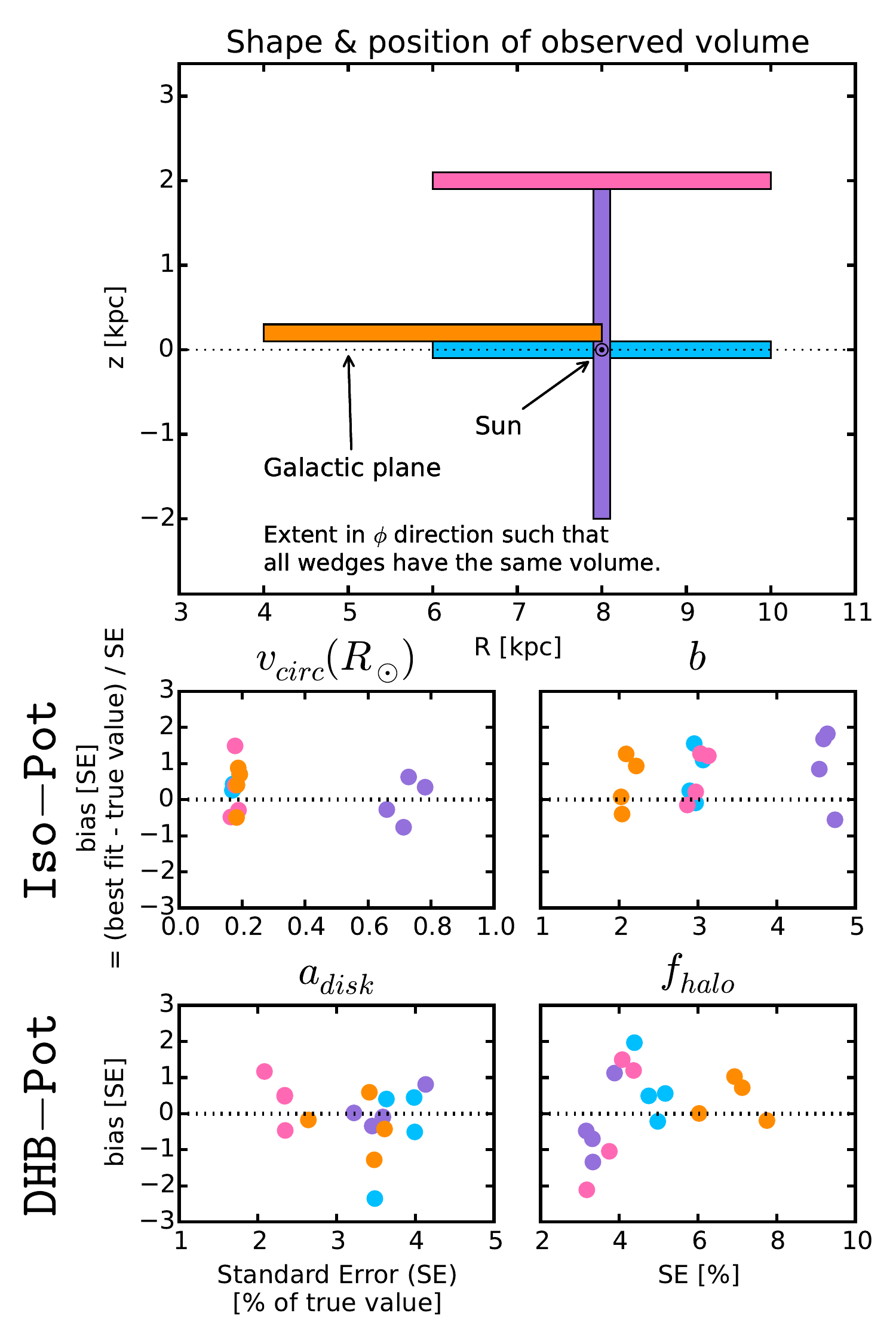}
\caption{Bias vs. standard error in recovering the potential parameters for mock data sets drawn from four different wedge-shaped test observation volumes within the Galaxy (illustrated in the upper panel; the corresponding analyses are colour-coded) and two different potentials (\texttt{Iso-Pot} and \texttt{DHB-Pot} from Table \ref{tbl:referencepotentials}; see also Test \ref{test:wedgesVol} in Table \ref{tbl:tests} for all model parameters used). Standard error and offset were determined from a Gaussian fit to the marginalized \pdf{}. The angular extent of each wedge-shaped observation volume was adapted such that all have a volume of $4.5~\text{kpc}^3$, even though their extent in $(R,z)$ is different. (The recovery of the free potential parameter $v_\text{circ}(R_\odot)$ in the different wedges is very similar for both potentials and therefore only shown for the \texttt{Iso-Pot}). Minor expected differences can be seen (e.g., $v_\text{circ}(R_\odot)$ and tracer density scale lengths requiring larger radial extent), but overall there is no clear trend that an observation volume around the Sun, above the disk or at smaller Galactocentric radii should give remarkably better constraints on the potential than the other volumes.}
\label{fig:wedgesVol_bias_vs_SE}
\end{figure}


\subsection{The role of the survey volume geometry} \label{sec:results_obsvolume}

To explore the role of the survey volume at given sample size, we devise two suites of mock data sets.

The first suite draws mock data for two different potentials (\texttt{Iso-Pot} and \texttt{DHB-Pot}) and four volume wedges (see Appendix \ref{app:selectionfunction}) with different extent and at different positions within the Galaxy, illustrated in the upper panel of Figure \ref{fig:wedgesVol_bias_vs_SE}. Otherwise the data sets are generated from the same \pmodel{} (see Test \ref{test:wedgesVol} in Table \ref{tbl:tests}). To isolate the role of the survey volume geometry and position, the mock data sets all have the same number of stars ($N_{*} = 20,000$), and are drawn from identical total survey volumes ($4.5~\text{kpc}^3$, achieved by adjusting the angular width of the wedges).
We investigate these rather unrealistic survey volumes to test (i) if there are regions in the Galaxy where stars are on intrinsically more informative orbits and (ii) if spatial cuts applied to the survey volume (e.g., to avoid regions of large dust extinction or measurement uncertainties) would therefore strongly affect the precision of the potential constraints. To make this effect---if it exists---noticeable, we choose some extreme, but illustrative examples.

The results are shown in Figure \ref{fig:wedgesVol_bias_vs_SE}: The wedges all have the same volume and all give results of similar precision. There are some minor and expected differences, e.g., $v_\text{circ}(R_\odot)$ and radial scale lengths ($b$ and $a_\text{disk}$) are slightly better recovered for large radial extent and the halo fraction at the Sun, $f_\text{halo}$, for volumes centered around $R_\odot$. In the case of an axisymmetric model galaxy, the extent in $\phi$ direction is not expected to matter. Overall radial extent and vertical extent seem to be equally important to constrain the potential. Figure \ref{fig:wedgesVol_bias_vs_SE} implies therefore that volume offsets or spatial cuts of the survey volume in the radial or vertical direction have at most a modest impact---even in case of the very large sample size at hand.

The second suite of mock data sets was already introduced in Section \ref{sec:largedata} (see also Test \ref{test:isoSph_CLT} in Table \ref{tbl:tests}), where mock data sets were drawn from five spherical volumes around the Sun with different $r_\text{max}$, for two different stellar populations. The results of this second suite are shown in Figure \ref{fig:isoSph_CLT} and exemplify the effect of the size of the survey volume.

Figure \ref{fig:isoSph_CLT} demonstrates that, given a choice of $p_\text{DF}$, a larger volume always results in tighter constraints. There is no obvious trend that a hotter or cooler population will always give better results; it depends on the survey volume and the model parameter in question.

While it appears that the argument for significant radial and vertical extent is generic, we have not done a full exploration of all combinations of \pmodel{} and volumes.

That in reality different regions in the Galaxy have different stellar number densities and different measurement uncertainties, should therefore be the major factor to drive the precision of the potential recovery when choosing a survey volume.

\begin{figure}[!htbp]
\centering
\includegraphics[width=\columnwidth]{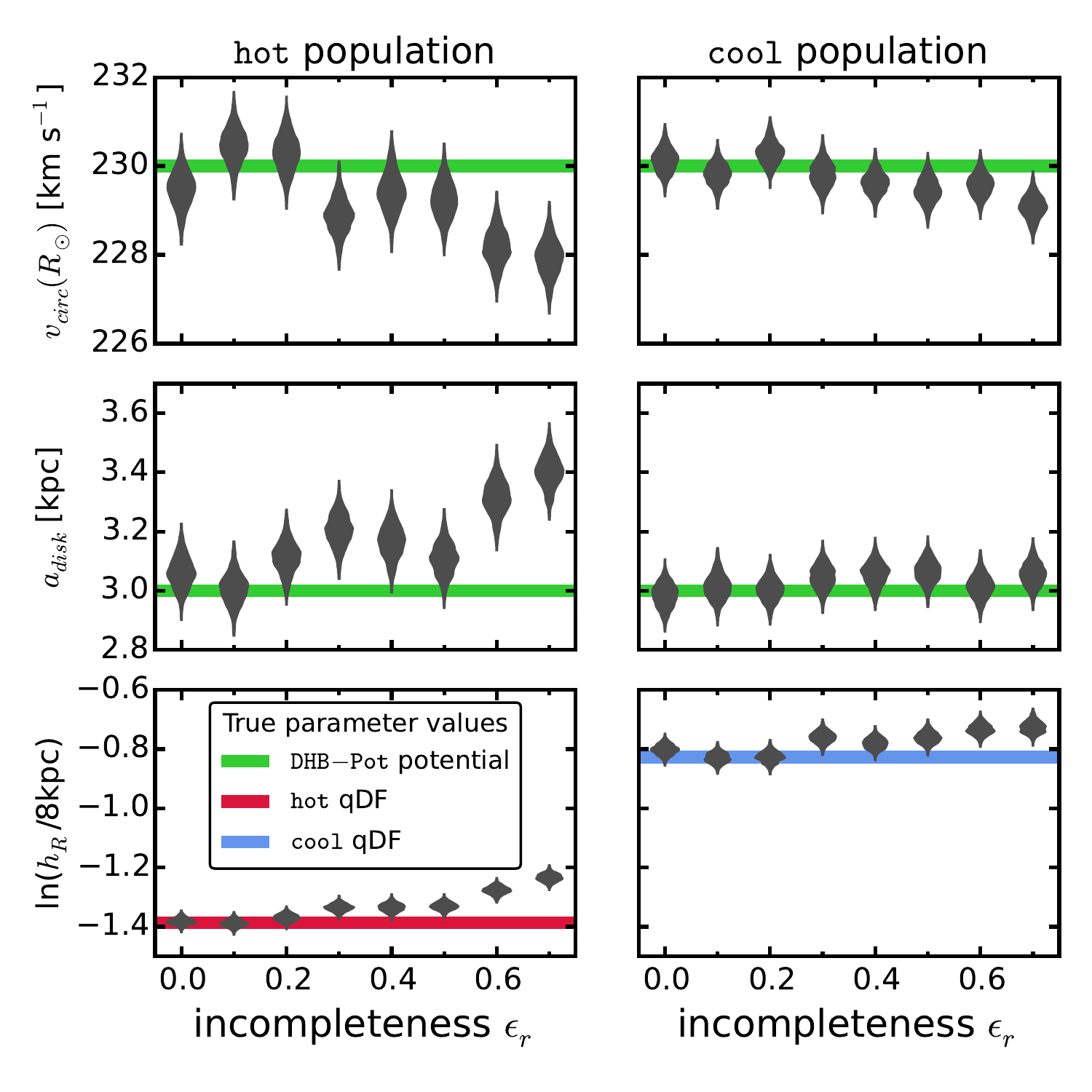}
\caption{Impact of misjudging the completeness of the data on the parameter recovery with \RM{}. Each mock data set was created with a different incompleteness parameter $\epsilon_r$ (shown on the $x$-axis, see Equation \eqref{eq:rad_incomp}). (The model parameters are given as Test \ref{test:isoSphFlexIncomp} in Table \ref{tbl:tests}.) The analysis however assumed that all data sets had constant completeness within the survey volume ($\epsilon_r = 0$). The violins show the full shape of the projected \pdf{}s for each model parameter, and the solid lines their true values. The \RM{} method seems to be robust against modest deviations between the true and the assumed data incompleteness. (The potential parameter $f_\text{halo}$ and the other qDF parameters are recovered to a comparable accuracy and are therefore not shown here.)} 
\label{fig:MWdhbIncompR_violins}
\end{figure}

\subsection{Impact of misjudging the selection function of the data set} \label{sec:results_incompR}

The survey SF (see also Appendix \ref{app:selectionfunction}) is sometimes not perfectly determined. While the pattern of the survey area on the sky may be complex, it is usually precisely known. It is the uncertainties (e.g., in completeness) in the line-of-sight directions that is most prone to systematic misassessment. We therefore focus here on completeness misjudgements (with a simplified angular pattern) and investigate how much this could affect the recovery of the potential. We do this by creating mock data in the \texttt{DHB-Pot} within a spherical survey volume with radius $r_\text{max}$ around the Sun (see Test \ref{test:isoSphFlexIncomp} in Table \ref{tbl:tests}) and a spatially varying completeness function
\begin{equation}
\text{completeness}(r) \equiv 1- \epsilon_r \frac{r}{r_\text{max}}, \label{eq:rad_incomp}
\end{equation}
which drops linearly with distance $r$ from the Sun. The completeness function can be understood as the probability of a star at distance $r$ to be detected (see also Equation \ref{eq:selectionfunction}). In the \RM{} analysis on the other hand, we assume constant completeness ($\epsilon_r=0$). The incompleteness parameter $\epsilon_r$ of the mock data quantifies therefore by how much we misjudge the SF. This mock test captures the relevant case of stars being less likely to be observed (than assumed) the further away they are (e.g., due to unknown dust obscuration). 

Figure \ref{fig:MWdhbIncompR_violins} demonstrates that the potential recovery with \RM{} is quite robust against somewhat wrong assumptions about the completeness of the data, i.e., $\epsilon_r \lesssim 0.15$ for the \texttt{hot} and $\epsilon_r \lesssim 0.2$ for the \texttt{cool} population. The \texttt{cool} population is more robust, because it is less affected by the SF misjudgement at high $|z|$ than the \texttt{hot} population. Our simple model SF affects stars at large and small radii in equal proportion. As long as the misjudgement is small, the tracer scale length parameter $h_R$ can still be reliable recovered, and with it the potential.

We have also investigated several test suites using the \texttt{Iso-Pot}. The recovery of $v_\text{circ}(R_\odot)$ and the qDF parameters at different $\epsilon_r$ is qualitatively and quantitatively similar to Figure \ref{fig:MWdhbIncompR_violins} for the \texttt{DHB-Pot}. The isochrone scale length $b$ however is recovered independently of $\epsilon_r$---probably because rotation curve measurements in the plane alone, which are not affected by the SF cuts, give reliable constraints on $b$. When not including tangential velocity measurements in the analysis (which is done by marginalizing the likelihood in Equation \eqref{eq:prob} over $v_T$), the parameters are well recovered only for $\epsilon_r \lesssim 0.15$ and $\epsilon_r \lesssim 0.2$ for the \texttt{hot} and \texttt{cool} population respectively. As this is in concordance with our findings for the $\texttt{DHB-Pot}$, this result seems to be valid for different choices of potentials.

For spatial completeness functions varying with the distance from the plane $|z|$ only, the $\texttt{Iso-Pot}$ potential recovery is similarly robust as long as $v_T$ measurements are included.

\begin{figure}[!htbp]
\centering
\includegraphics[width=\columnwidth]{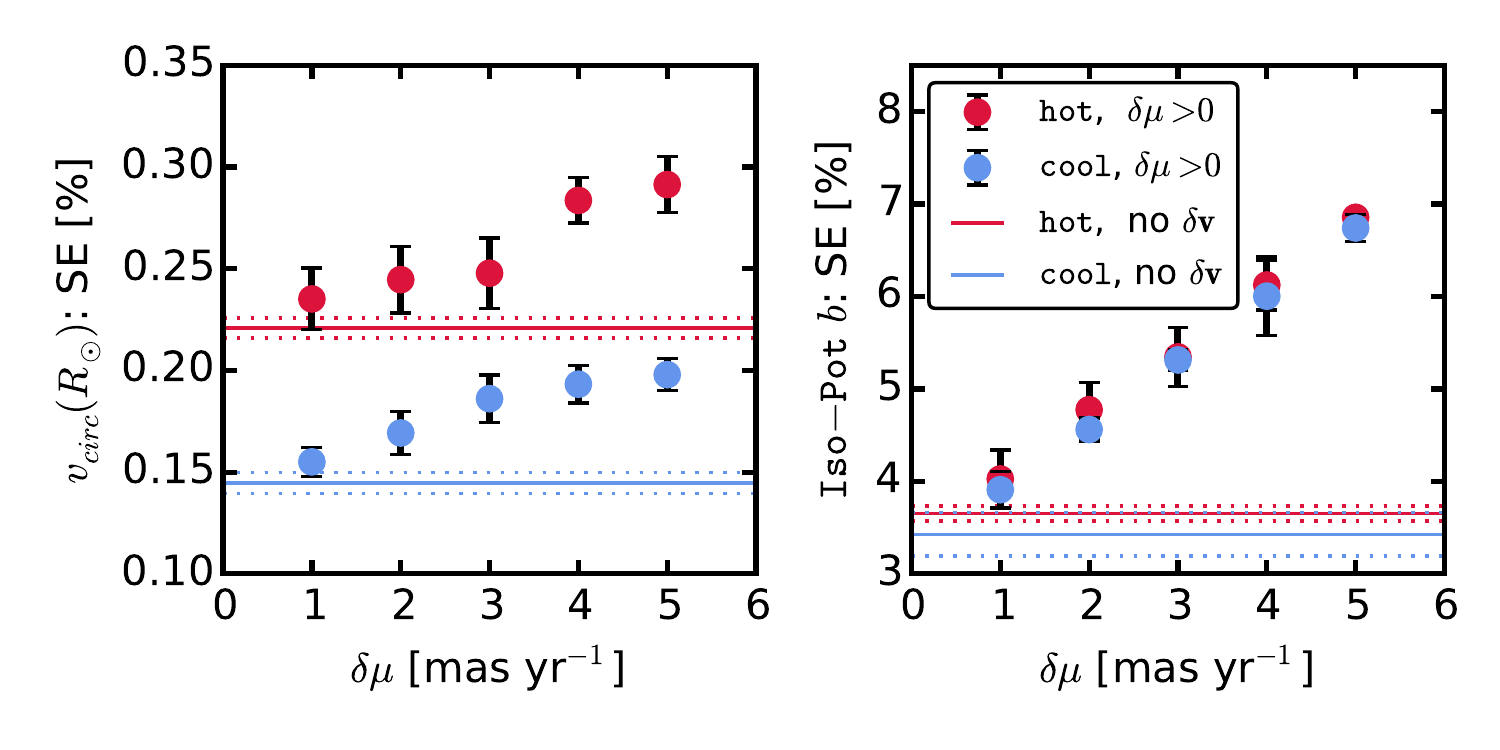}
\caption{Effect of proper motion uncertainties $\delta \mu$ on the precision of potential parameter recovery for two stellar populations of different kinematic temperature (see Test \ref{test:isoSphFlexErrConv_SE_vs_error} in Table \ref{tbl:tests} for all model parameters). The relative standard error (SE) derived from the marginalized \pdf{} for each model parameter was determined for precise data sets without measurement uncertainties (solid lines, with dotted lines indicating the error) and for data sets affected by different proper motion uncertainties $\delta \mu$ and $\delta v_\text{los}=2~\text{km s}^{-1}$ (data points with error bars), but no uncertainties in position. The errors come from taking the mean over several data sets.}
\label{fig:isoSphFlexErrConv_SE_vs_error}
\end{figure}

\begin{figure}[!htbp]
\centering
\includegraphics[width=\columnwidth]{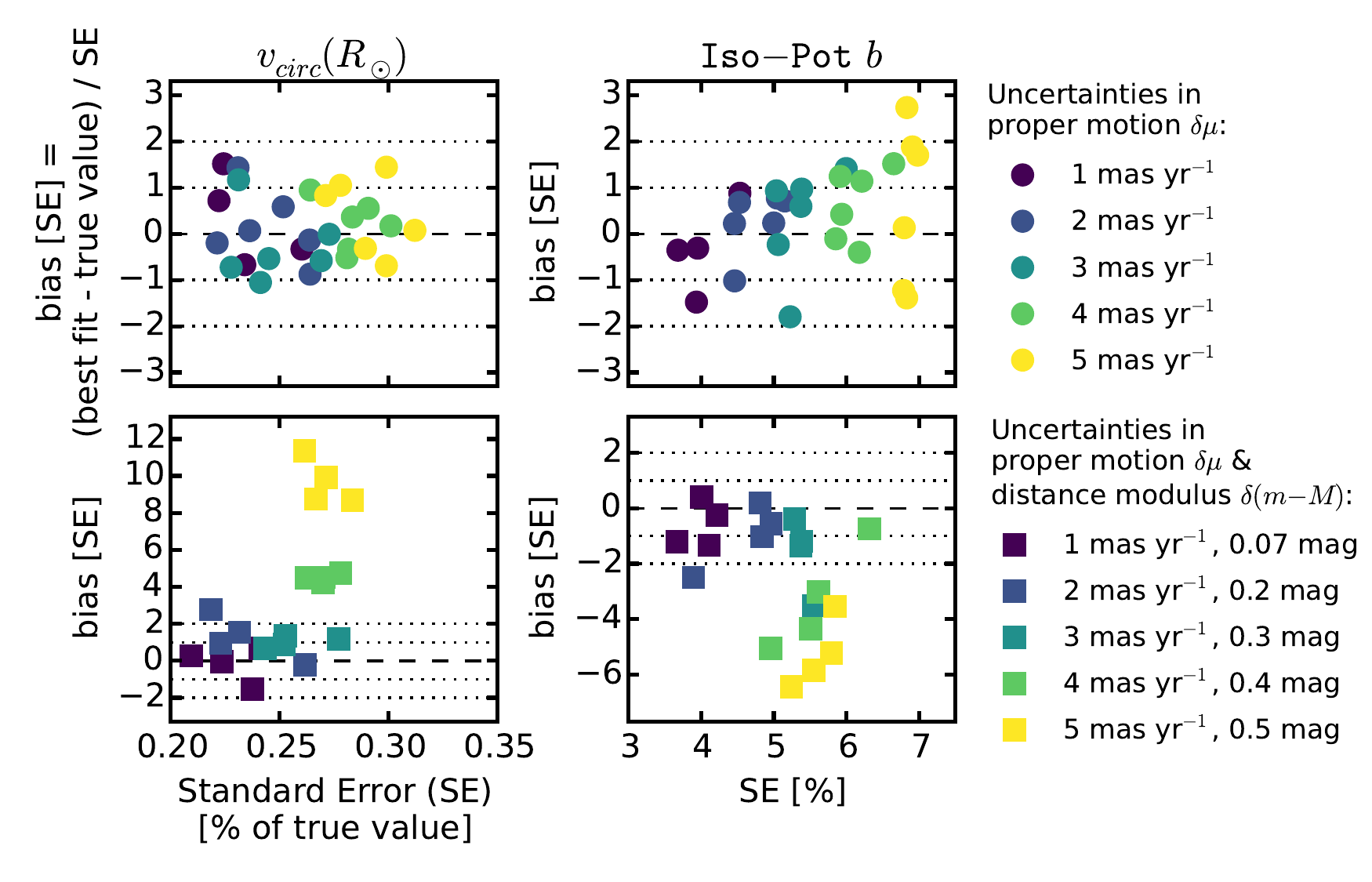}
\caption{Potential parameter recovery using the approximation for the model probability convolved with measurement uncertainties in Equation \eqref{eq:errorconv}. We show \pdf{} offset and relative width (i.e., standard error SE) for potential parameters recovered from mock data sets (which were created according to Test \ref{test:isoSphFlexErrConv_bias_vs_SE} in Table \ref{tbl:tests}). The data sets in the upper panels are affected only by proper motion uncertainties $\delta \mu$ (and $\delta v_\text{los}=2~\text{mas yr}^{-1}$), while the data sets in the lower panels also have distance (modulus) uncertainties $\delta (m-M)$, as indicated in the legend. For data sets with $\delta \mu \leq 3 ~\text{mas yr}^{-1}$ Equation \eqref{eq:errorconv} was evaluated with $N_\text{samples}=800$, for $\delta \mu > 3~\text{mas yr}^{-1}$ we used $N_\text{samples}=1200$. In absence of distance uncertainties Equation \eqref{eq:errorconv} gives unbiased results. For $\delta(m-M) > 0.2~\text{mag}$ (i.e., $\delta r/r > 0.1$; for $r \sim 3~\text{kpc}$) however biases of several $\sigma$ are introduced, as Equation \eqref{eq:errorconv} is only an approximation for the true likelihood in this case.}
\label{fig:isoSphFlexErrConv_bias_vs_SE}
\end{figure}

\begin{figure}[!htbp]
\centering
\includegraphics[width=\columnwidth]{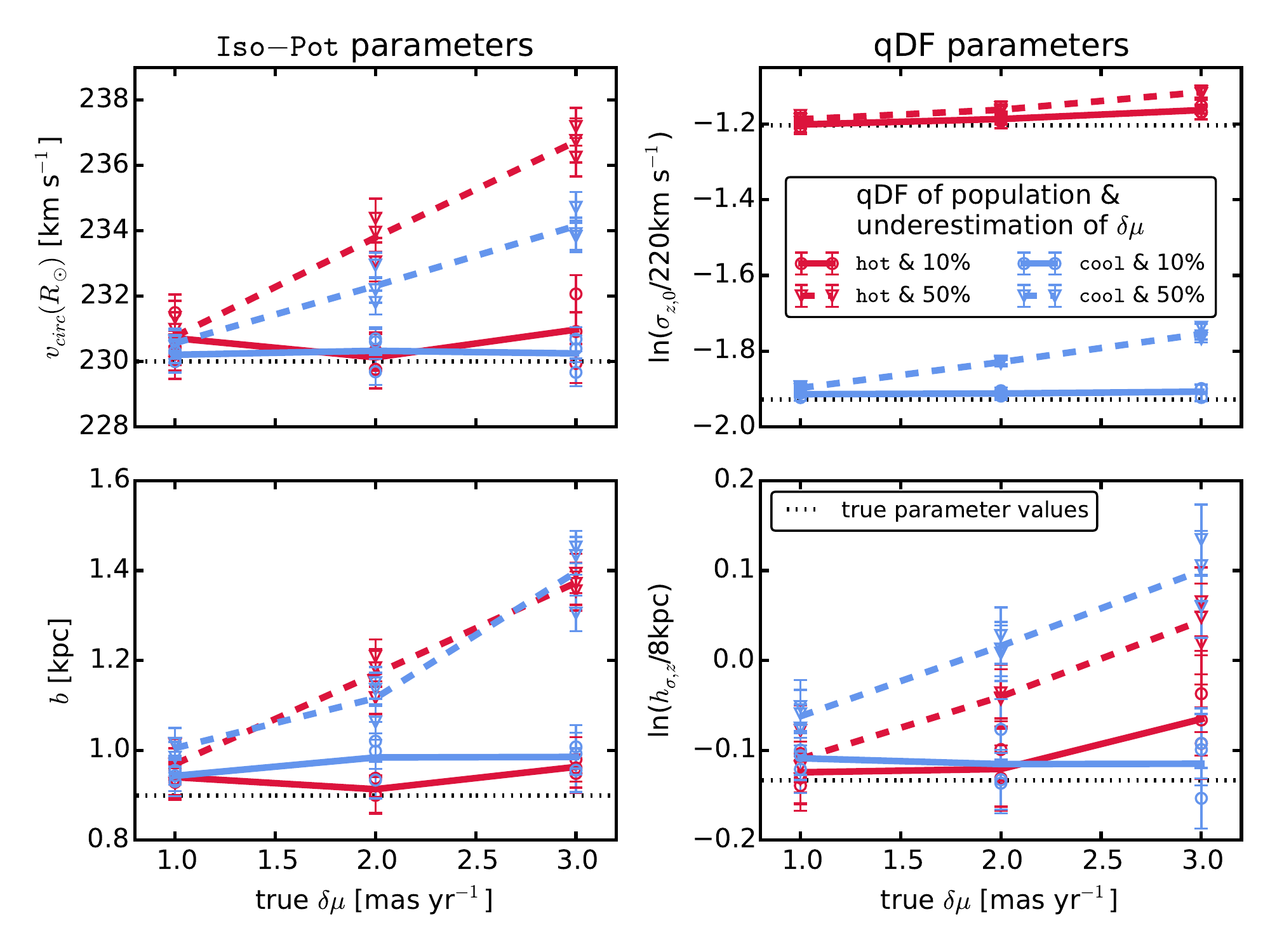}
\caption{Effect of a systematic underestimation of proper motion uncertainties $\delta \mu$ on the recovery of the model parameters. (The true model parameters used to create the mock data are summarized as Test \ref{test:isoSphFlexErrSyst} in Table \ref{tbl:tests}, four of them are indicated as black dotted lines in this figure.) The mock data was perturbed according to proper motion uncertainties $\delta \mu = \delta \mu_\text{Dec} = \delta \mu_\text{RA}$ as indicated on the $x$-axis. In the \RM{} analysis (see likelihood in Equation \eqref{eq:errorconv}) however, we underestimated the true $\delta \mu$ by 10\% (circles, solid lines) and 50\% (triangles, dashed lines). The symbols denote the best fit parameters with $1\sigma$ error bars of several mock data sets. The lines connect the mean of corresponding data realisations to guide the eye.}
\label{fig:isoSphFlexErrSyst}
\end{figure}

\subsection{Measurement uncertainties and their effect on the parameter recovery} \label{sec:results_errors}

Measurement uncertainties in proper motions and distance dominate over uncertainties in position on the sky (RA, Dec) and line-of-sight velocity, which can be more accurately determined.

The range of proper motion uncertainties we will investigate in this section, $1-5~\text{mas yr}^{-1}$, is the approximate measurement accuracy that can be achieved by combining catalogues from ground-based surveys like the Sloan Digital Sky Survey (SDSS; \citealt{2003AJ....126.2081A}), the USNO-B catalogue \citep{2003AJ....125..984M}, 2MASS \citep{2006AJ....131.1163S} and the Pan-STARRS1 photometric catalogue (PS1; \citealt{2010SPIE.7733E..0EK}).\footnote{Combining observations from the SDSS Data Release 1 with the USNO-B catalogue based on the Palomar Observatory Sky Survey's (POSS) photographic plates from the 1950s lead to proper motion measurements precise to $\delta\mu\sim3$ or $5~\text{mas yr}^{-1}$ depending on magnitude $r<18$ or $r<20$ respectively \citep{2004AJ....127.3034M,2008AJ....136..895M,2004ApJS..152..103G}. The same accuracy can be achieved when using the four years of measurements by the PS1 only. By careful calibration of USNO-B and 2MASS with PS1, \citet{2015ApJ...809...59S} even got proper motions as accurate as $\delta \mu\sim1.5~\text{mas yr}^{-1}$ for $r\lesssim 18$. The Large Synoptic Survey Telescope (LSST, \citealt{2008arXiv0805.2366I}) planned for 2021 might even achieve $\delta\mu \lesssim 1~\text{mas yr}^{-1}$ during its 10 years of scanning the sky \citep{2008IAUS..248..537I}.} Space-based surveys can do even better: The Hipparcos \citep{1997ESASP1200.....E} and Tycho-2 \citep{2000A&A...355L..27H} catalogues achieve $\delta \mu\sim2.5~\text{mas yr}^{-1}$ (and even $\delta\mu \lesssim 1~\text{mas yr}^{-1}$ for all stars with $V <12$), which will be soon superseded by Gaia with only $\delta \mu\sim0.3~\text{mas yr}^{-1}$ at its faint end at magnitude $G\sim20$ \citep{2014EAS....67...23D}.

We first investigate the impact of (perfectly known) proper motion uncertainties on the precision of the potential parameter recovery (see Test \ref{test:isoSphFlexErrConv_SE_vs_error} in Table \ref{tbl:tests}). Figure \ref{fig:isoSphFlexErrConv_SE_vs_error} demonstrates that for data sets with $\delta \mu$ as high as $5~\text{mas yr}^{-1}$ the precision degrades by a factor of no more than $\sim2$ as compared to a data set without measurement uncertainties. The precision gets monotonically better for smaller $\delta \mu$, being larger only by a factor of $\sim 1.15$ at $\delta \mu=1~\text{mas yr}^{-1}$. With relative standard errors on the recovered parameters of only a few percent at most for 10,000 stars, this means we still get quite precise constraints on the potential, as long as we know the proper motion uncertainties perfectly.

We also note that in this case the relative and absolute difference in recovered precision between the precise and the uncertainty-affected data sets does not seem to depend strongly on the kinematic temperature of the stellar population.

Secondly, we investigate the impact of additional measurement uncertainties in distance (modulus). In absence of distance uncertainties the uncertainty-convolved model probability given in Equation \eqref{eq:errorconv} is unbiased (see upper left panel in Figure \ref{fig:isoSphFlexErrConv_bias_vs_SE}). When including distance (modulus) uncertainties, Equation \eqref{eq:errorconv} is just an approximation for the true likelihood; the systematic bias thus introduced in the parameter recovery gets larger with the size of $\delta (m-M)$, as demonstrated in Figure \ref{fig:isoSphFlexErrConv_bias_vs_SE}, lower panels (see also Test \ref{test:isoSphFlexErrConv_bias_vs_SE} in Table \ref{tbl:tests}). We find however that in case of $\delta(m-M) \lesssim 0.2 \text{ mag}$ (if also $\delta \mu \lesssim 2 ~\text{mas yr}^{-1}$ and a maximum distance of $r_\text{max} = 3~\text{kpc}$, see Test \ref{test:isoSphFlexErrConv_bias_vs_SE} in Table \ref{tbl:tests}) the potential parameters can still be recovered within $2 \sigma$. This corresponds to a relative distance uncertainty of $\sim10\%$. The overall precision of the potential recovery is also not degraded much by introducing distance uncertainties of less than $10\%$.

How does this compare with the distance uncertainties expected for Gaia? For a typical red clump giant star with $M_I\sim0~\text{mag}$ and $(V-I)\sim1~\text{mag}$ at a distance of $r=3~\text{kpc}$ we estimate (using the magnitude transformation by \citet{2010A&A...523A..48J} and the uncertainty parametrization by \citet{2014EAS....67...23D}) a parallax uncertainty of $\delta\pi\sim11~\mu\text{as}$, which is consistent with a distance uncertainty of less than 5\%, and a proper motion uncertainty of $\delta \mu\sim6~\mu\text{as yr}^{-1}$, which is negligible. For the case that the modelling is not restricted to giant stars only, a quick investigation by Rene Andrae (private communication) of stars at $r=3~\text{kpc}\pm5~\text{pc}$ from the Gaia Universe model snapshot catalogue (GUMS; \citealt{2012A&A...543A.100R}) revealed that a magnitude cut at $G\sim15~\text{mag}$ in the overall Gaia data set should keep all distance uncertainties within $3~\text{kpc}$ below $\sim10\%$, while also preserving Gaia's simple SF.

We therefore found that in case we perfectly know the measurement uncertainties (and the distance uncertainty is negligible or of the order of the uncertainties expected from Gaia within $\sim3~\text{kpc}$), the convolution of the model probability with the measurement uncertainties gives \emph{precise and accurate} constraints on the model parameters---even if the measurement uncertainty itself is quite large.

\begin{figure}[!htbp]
\centering
\includegraphics[width=\columnwidth]{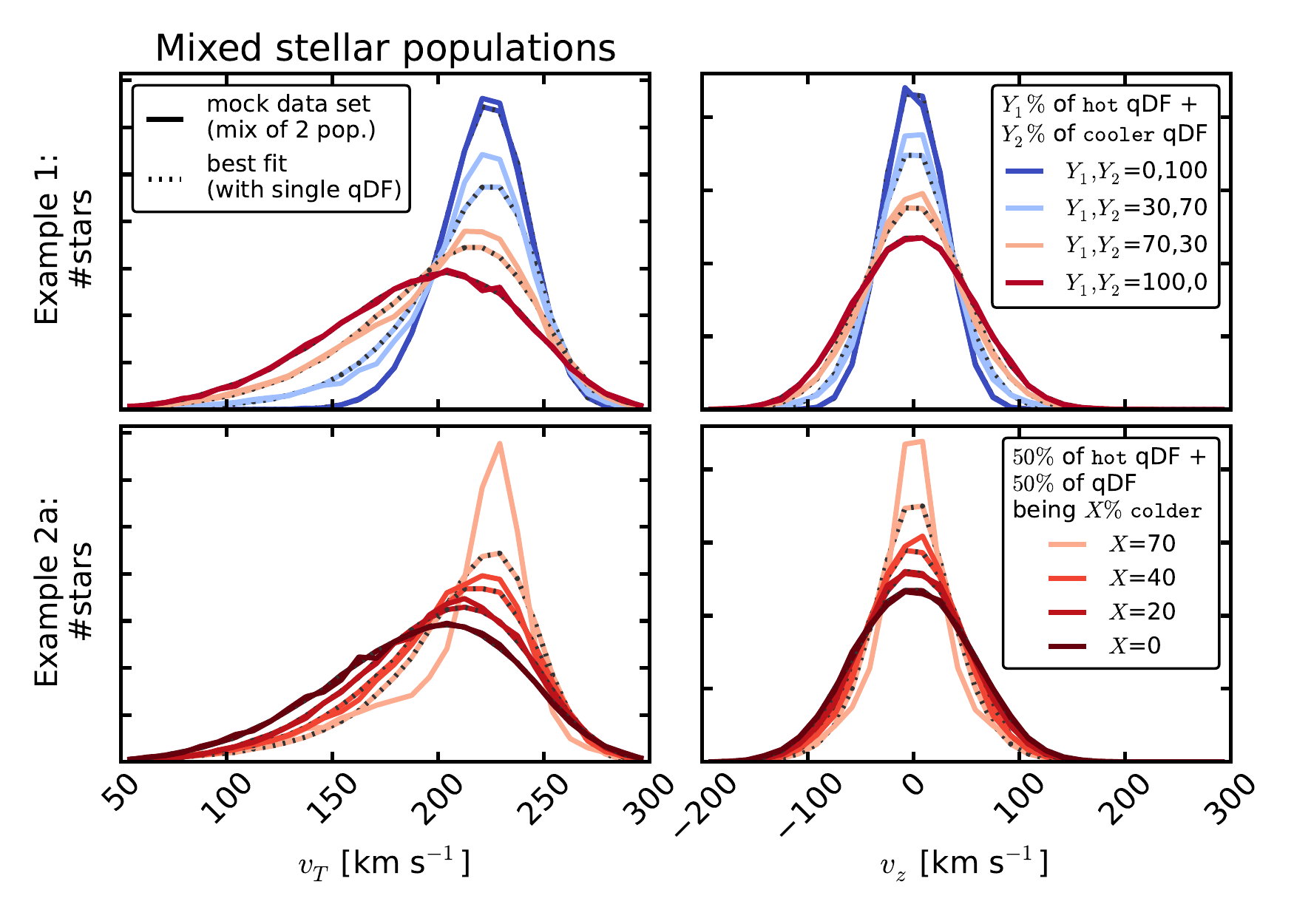}
\caption{Distribution of mock data $v_T$ and $v_z$ created by mixing stars drawn from two different qDFs (solid lines), and the distribution predicted by the best fit of a single qDF and potential to the data (dotted lines). (The model parameters used to create the mock data are given in Table \ref{tbl:tests} as Test \ref{test:MWdhbMix}, \emph{Example 1 \& 2a}, with the qDF parameters referred to in the legend given in Table \ref{tbl:referenceMAPs}.) The corresponding single qDF best fit curves were derived from the best fit parameters found in Figures \ref{fig:MWdhbMixCont} and \ref{fig:MWdhbMixDiff}. (The data sets are colour-coded in the same way as the corresponding analyses in Figures \ref{fig:MWdhbMixCont} and \ref{fig:MWdhbMixDiff}.) We use the mixtures of two qDFs to demonstrate how \RM{} behaves for data sets following DFs with shapes slightly differing from a single qDF. For large deviations it might already become visible from directly comparing the mock data and best fit distribution, that a single qDF is a bad assumption for the stars' true DF.}
\label{fig:MWdhbMix_mockdata_residuals}
\end{figure}

\begin{figure}[!htbp]
\centering
\includegraphics[width=\columnwidth]{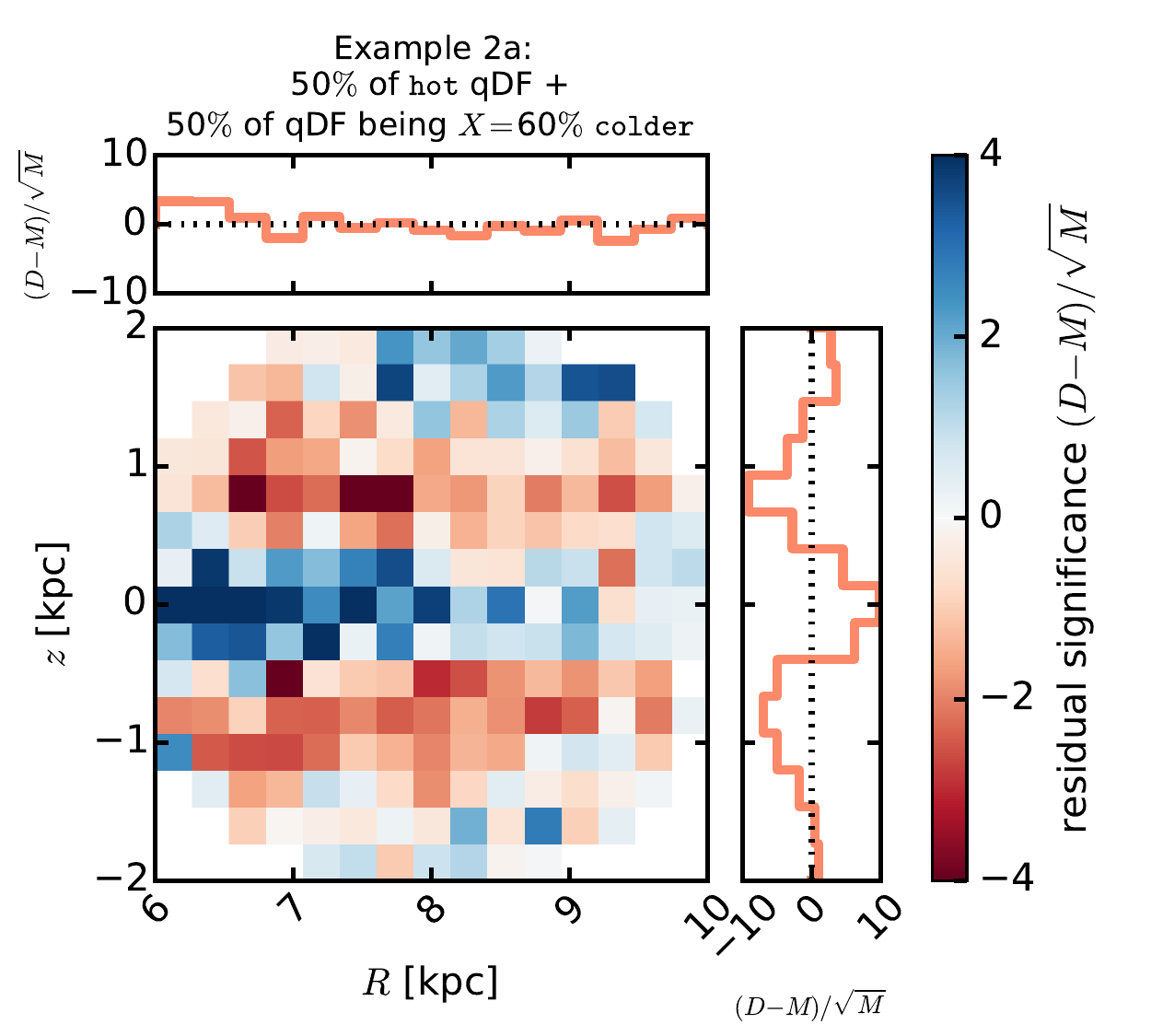}
\caption{Residual significance $(D-M)/\sqrt{M}$ of one example mock data set $D$ and its best fit single qDF model $M$ in the $(R,z)$ plane. The mock data set $D$ was created by mixing a \texttt{hot} and a $X\%$ \texttt{colder} population in equal proportion (see also Table \ref{tbl:tests}, Test \ref{test:MWdhbMix}, \emph{Example 2a}, with $X=60\%$). The best fit distribution $M$ was derived analogously to the ones shown for the velocity components in Figure \ref{fig:MWdhbMix_mockdata_residuals}. This is an extreme example where the best fit single qDF is not a good fit anymore (see Figure \ref{fig:MWdhbMixDiff}, \emph{Example 2a}, $X=60\%$), but it illustrates how we constructed mock data distributions with radial and vertical density profiles differing from a single qDF by mixing two different qDF populations for the Test suite \ref{test:MWdhbMix}.}
\label{fig:MWbdhMix_residual_RvsZ}
\end{figure}

\begin{figure}[!htbp]
\centering
\includegraphics[scale=0.55]{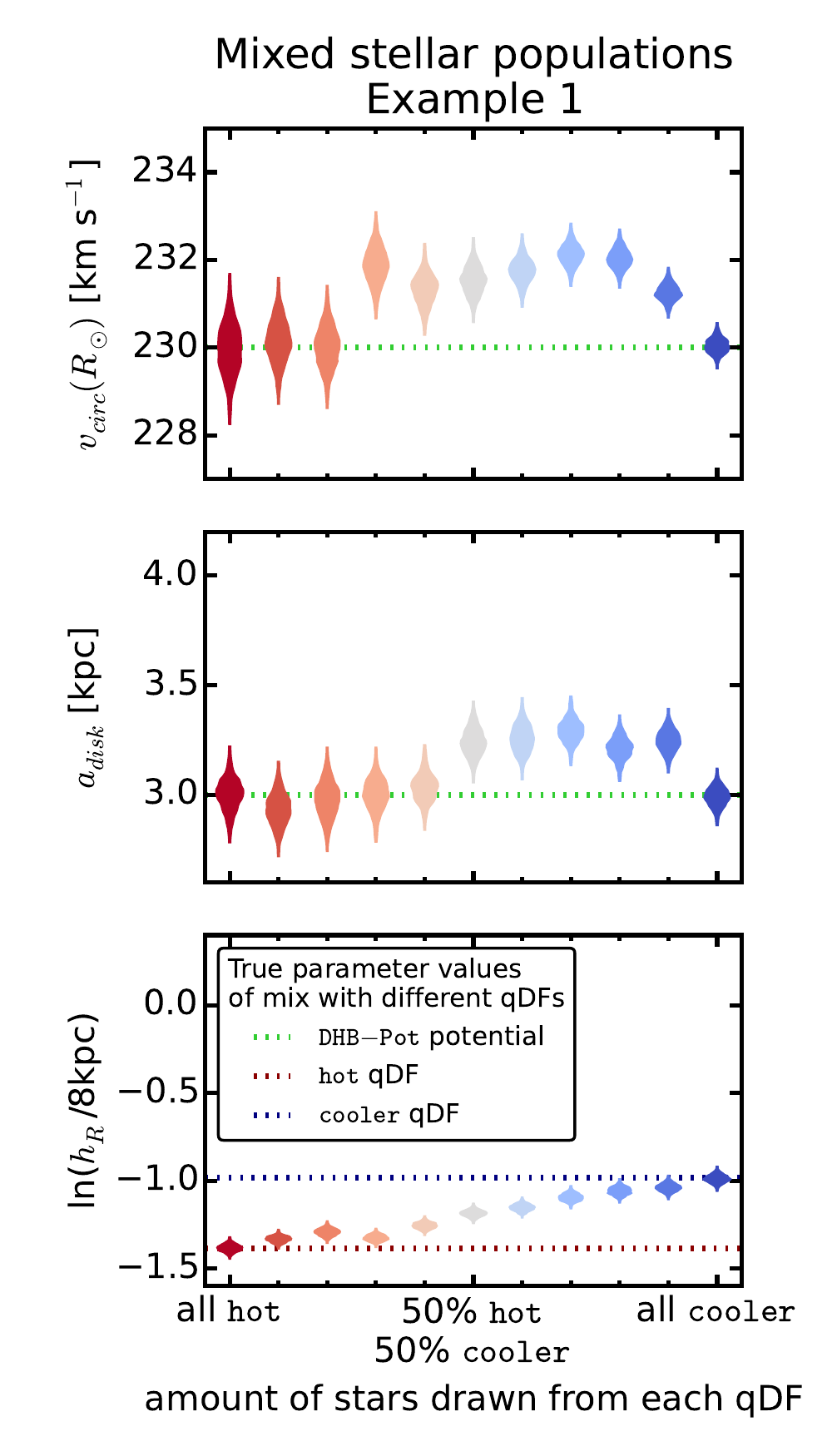}
\caption{The dependence of the parameter recovery on degree of pollution and temperature of the stellar population. We mix (i.e., ``pollute'') varying amounts of stars from a \texttt{hot} stellar population with stars from a very different \texttt{cooler} population (see Table \ref{tbl:referenceMAPs}), as indicated on the $x$-axis. (All model parameters used to create the mock data are given as Test \ref{test:MWdhbMix}, \emph{Example 1}, in Table \ref{tbl:tests}.) The composite polluted mock data set follows a true DF that has a slightly different shape than the qDF. We then analyse it using \RM{} and fit a \emph{single} qDF only. The violins represent the marginalized \pdf{}s for the best fit model parameters. Some mock data sets are shown in Figure \ref{fig:MWdhbMix_mockdata_residuals}, first row, in the same colours as the violins here. We find that a hot population is much less affected by pollution with stars from a cooler population than vice versa. (The potential parameter $f_\text{halo}$ is recovered to a similar or even slightly better accuracy than $a_\text{disk}$ at each given mixing rate and is therefore not shown here.)}
\label{fig:MWdhbMixCont}
\end{figure}

\begin{figure}[!htbp]
\centering
\includegraphics[scale=0.55]{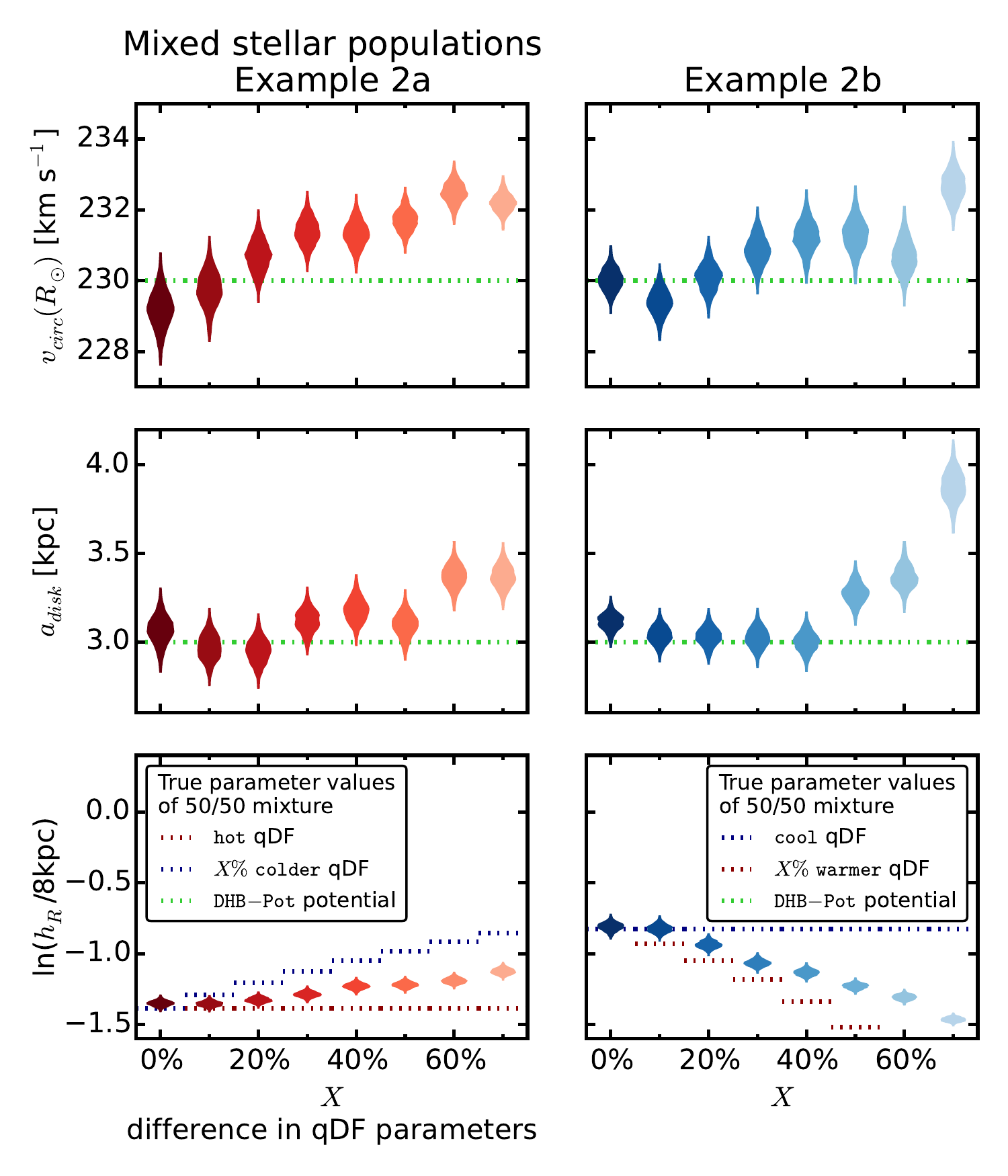}
\caption{The dependence of the parameter recovery on the difference in qDF parameters of a 50/50 mixture of two stellar populations and their temperature. The two qDFs from which the stars in each mock data set were drawn are indicated in the legend, with the qDF parameters $\sigma_{R,0}, \sigma_{z,0}$ and $h_R$ differing by $X\%$ (see also Table \ref{tbl:referenceMAPs} and Section \ref{sec:qDF}), as indicated on the $x$-axis. (The model parameters used for the mock data creation are given as Test \ref{test:MWdhbMix}, \emph{Example 2a \& b}, in Table \ref{tbl:tests}.) Each composite mock data set is fitted with a \emph{single} qDF and the marginalized \pdf{}s are shown as violins. Some mock data sets of \emph{Example 2a} and their best fit distributions are shown in Figure \ref{fig:MWdhbMix_mockdata_residuals}, last row (colour-coded analogous to the violins here), and Figure \ref{fig:MWbdhMix_residual_RvsZ} shows the corresponding residuals in the $(R,z)$ plane. By mixing populations with varying difference in their qDF parameters, we model the effect of finite bin size or abundance errors when sorting stars into different \MAPs{} in the $[\alpha/\mathrm{Fe}]$-vs.-$[\mathrm{Fe}/\mathrm{H}]$ plane and assuming they follow single qDFs (cf. BR13). We find that the bin sizes should be chosen such that the difference in qDF parameters between neighbouring \MAPs{} is less than 20\%. (The potential parameter $f_\text{halo}$ is recovered to a similar or even slightly better accuracy than $a_\text{disk}$ at each given $X$ and is therefore not shown here.)} 
\label{fig:MWdhbMixDiff}
\end{figure}

Lastly, Figure \ref{fig:isoSphFlexErrSyst} investigates the effect of a systematic underestimation of the true proper motion uncertainties $\delta \mu$ by 10\% and 50\% (see also Test \ref{test:isoSphFlexErrSyst} in Table \ref{tbl:tests}). We find that this causes a bias in the parameter recovery that grows seemingly linear with $\delta \mu$. For an underestimation of only $10\%$ however, the bias becomes $\lesssim 2 \sigma$ for 10,000 stars---even for $\delta \mu \sim 3~\text{mas yr}^{-1}$.

The size of the bias also depends on the kinematic temperature of the stellar population and the model parameter considered (see Figure \ref{fig:isoSphFlexErrSyst}). The qDF parameters are for example better recovered by hotter populations. This is, because the relative difference between the \emph{true} $\sigma_i(R)$ (with $i \in \{R,z\}$) and \emph{measured} $\sigma_i(R)$ (which comes from the deconvolution with an underestimated velocity uncertainty) is smaller for hotter populations. 

\subsection{The impact of deviations of the data from the idealized distribution function} \label{sec:results_mixedDFs}

Our modelling approach assumes that each stellar population follows a simple DF; here we use the qDF. In this section we explore what happens if this idealization does not hold. We investigate this issue by creating mock data sets that are drawn from \emph{two} distinct qDFs of different temperature\footnote{Following the observational evidence, our mock data populations with cooler qDFs also have longer tracer scale lengths.} (see Table \ref{tbl:referenceMAPs} and Test \ref{test:MWdhbMix} in Table \ref{tbl:tests}) in the \texttt{DHB-Pot}, and analyse the composite mock data set by fitting a \emph{single} qDF to it. The velocity distribution of some mock data sets and their best fit qDFs are illustrated in Figure \ref{fig:MWdhbMix_mockdata_residuals}, and Figure \ref{fig:MWbdhMix_residual_RvsZ} shows the tracer density residuals between data and best fit in the $(R,z)$ plane. Figures \ref{fig:MWdhbMixCont} and \ref{fig:MWdhbMixDiff} compare the input and best fit parameters. In \emph{Example 1} we choose qDFs of widely different temperature and vary their relative fraction of stars in the composite mock data set (Figure \ref{fig:MWdhbMixCont}); in \emph{Example 2} we always mix mock data stars from two different qDFs in equal proportion, but vary by how much the qDFs' temperatures differ (Figure \ref{fig:MWdhbMixDiff}). 

The first set of tests mimics a DF that has wider wings or a sharper core in velocity space than a qDF (see Figure \ref{fig:MWdhbMix_mockdata_residuals}) and slightly different radial and vertical tracer density profiles (similar to Figure \ref{fig:MWbdhMix_residual_RvsZ}). The second test could be understood as mixing neighbouring \MAPs{} in the $[\alpha/\mathrm{Fe}]$-vs.-$[\mathrm{Fe}/\mathrm{H}]$ plane due to large bin sizes or abundance measurement errors (cf. BR13). 

We consider the impact of the DF deviations on the recovery of the potential and the qDF parameters separately. 

We find from \emph{Example 1} that the potential parameters can be more robustly recovered, if a mock data population is polluted by a modest fraction ($\lesssim 30\%$) of stars drawn from a much cooler qDF, as opposed to the same pollution of stars from a hotter qDF. When considering the case of a 50/50 mix of contributions from different qDFs in \emph{Example 2}, there is a systematic, but mostly small, bias in recovering the potential parameters, monotonically increasing with the qDF parameter difference. In particular for fractional differences in the qDF parameters of $\lesssim 20\%$ the systematics are insignificant even for sample sizes of $N_{*} = 20,000$, as used in the mock data.

Overall, the circular velocity at the Sun is very reliably recovered to within $2\%$ in all these tests. But the best fit $v_\text{circ}(R_\odot)$ is not always unbiased at the implied precision.

The recovery of the effective qDF parameters, in light of non-qDF mock data, is quite intuitive (in Figures \ref{fig:MWdhbMixCont} and \ref{fig:MWdhbMixDiff} we therefore show only $h_R$): the effective qDF temperature lies between the two temperatures from which the mixed DF of the mock data was drawn; in all cases the scale lengths of the velocity dispersion fall-off, $h_{\sigma,R}$ and $h_{\sigma,z}$, are shorter than the true scale lengths, because the stars drawn form the hotter qDF dominate at small radii, while stars from the cooler qDF (with its longer tracer scale length) dominate at large radii; the recovered tracer scale lengths, $h_R$, vary smoothly between the input values of the two qDFs that entered the mix of mock data. The latter is also demonstrated in Figure \ref{fig:MWbdhMix_residual_RvsZ}: The radial tracer density profile of the mock data is steeper than a single qDF in the mid-plane and more shallow at higher $|z|$; overall the best fit $h_R$ lies therefore in between.

We note that in the cases where the systematic bias in the potential parameter recovery becomes several $\sigma$ large, a direct comparison of the true mock data set and best fit distribution (see Figure \ref{fig:MWdhbMix_mockdata_residuals}) can sometimes already reveal that the assumed DF is not a good model for the data.

We performed the same tests also for the spherical \texttt{Iso-Pot} instead of the galaxy-like \texttt{DHB-Pot} and for a much higher sampling of the mixing rate and qDF difference $X$. The results are qualitatively and quantitatively very similar and therefore independent of the exact choice of potential.

Overall, we find that the potential inference is quite robust to modest deviations of the data from the assumed DF. 


\subsection{The implications of a gravitational potential not from the space of model potentials} \label{sec:results_potential}

We now explore what happens when the mock data were drawn from one axisymmetric potential family, here \texttt{MW14-Pot}, and is then modelled considering potentials from another axisymmetric family, here \texttt{KKS-Pot} (see Table \ref{tbl:referencepotentials} and Figure \ref{fig:ref_pots}). In the analysis we assume the circular velocity at the Sun to be fixed and known and only fit the parametric potential form.\footnote{We made sure that $v_\text{circ}(R_\odot)$ can be very well recovered when included in the fit of a \texttt{cool} population. The model assumption that $v_\text{circ}(R_\odot)$ is known does therefore not affect the discussion qualitatively.}

\begin{figure}[!htbp]
\centering
\includegraphics[width=\columnwidth]{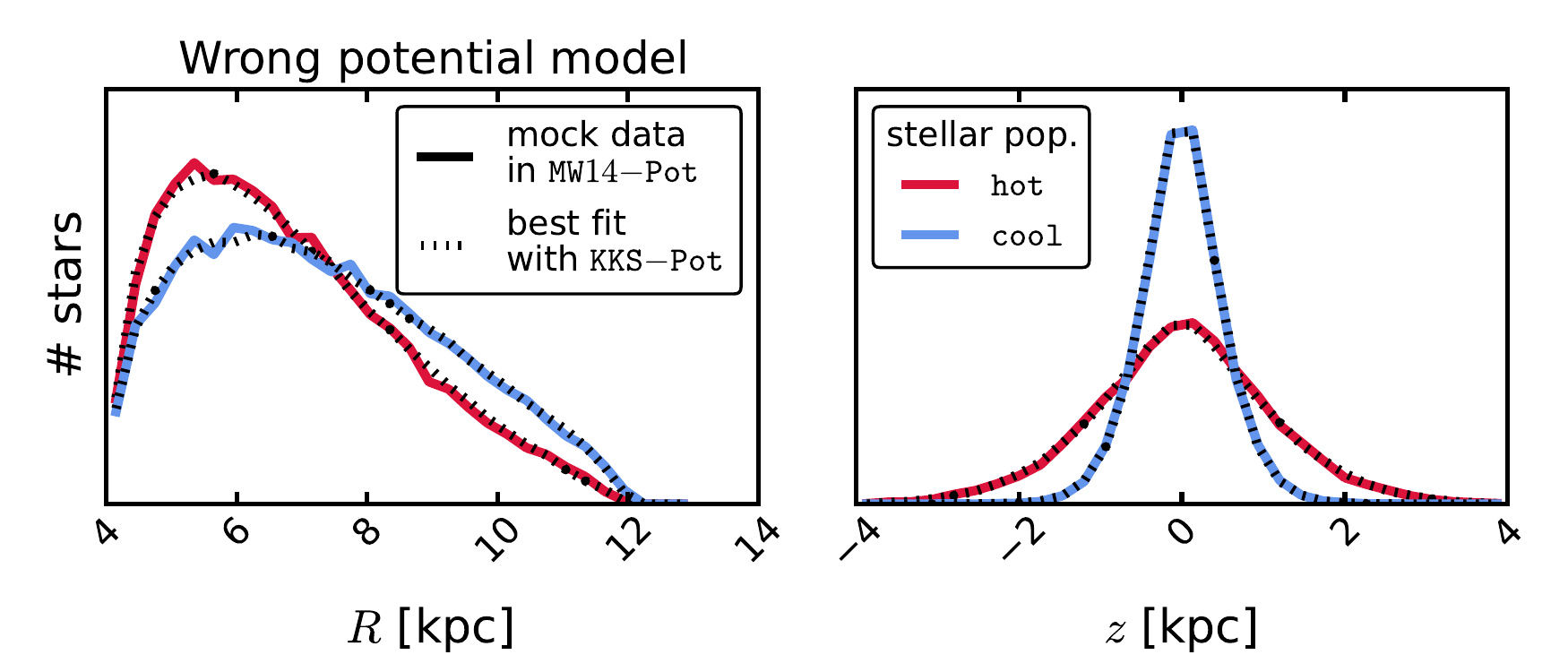}
\caption{Comparison of the spatial distribution of mock data in $R$ and $z$ created in the \texttt{MW14-Pot} potential and with two different stellar populations (see Test \ref{test:MW14vsKKS2New} in Table \ref{tbl:tests} for all mock data model parameters), and the best fit distribution recovered by fitting the family of \texttt{KKS-Pot} potentials to the data. The best fit potentials are shown in Figure \ref{fig:MW14vsKKS2New_contours} and the corresponding best fit qDF parameters in Figure \ref{fig:MW14vsKKS2New_violins}. The data is very well recovered, even though the fitted potential family did not incorporate the true potential.}
\label{fig:MW14vsKKS2New_mockdata_residuals}
\end{figure}

We analyse a mock data set from a \texttt{hot} and \texttt{cool} stellar population each (see Test \ref{test:MW14vsKKS2New} in Table \ref{tbl:tests}) with high numerical accuracy. The distributions generated from the best fit parameters reproduce the data in configuration space very well (see Figure \ref{fig:MW14vsKKS2New_mockdata_residuals} for the spatial distribution and the circles in Figure \ref{fig:MW14vsKKS2New_violins} for the velocity distribution).

\begin{figure*}[!htb]
\centering
\includegraphics[width=1\textwidth]{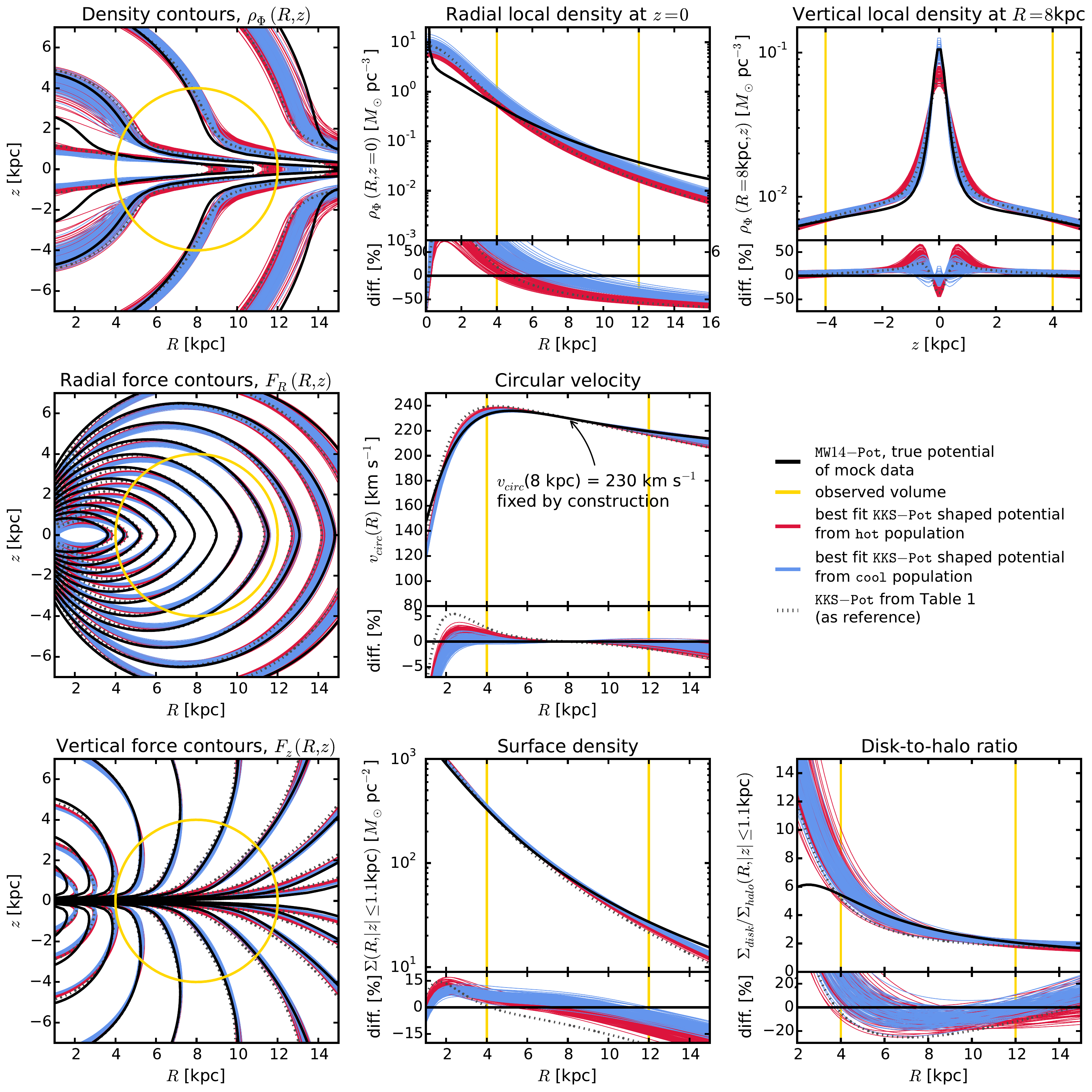}
\caption{Recovery of the gravitational potential if the assumed potential model family (\texttt{KKS-Pot} with fixed $v_\text{circ}(R_\odot)$) and the true potential of the (mock data) stars (\texttt{MW14-Pot} in Table \ref{tbl:referencepotentials}) have slightly different parametric forms. In addition to contours of equal density $\rho_\Phi$, radial and vertical force $F_R$ and $F_z$ in the $(R,z)$ plane (left comlumn), we show local density profiles $\rho_\Phi(R,z=0)$ and $\rho_\Phi(R=8~\text{kpc},z)$, as well as the circular velocity curve $v_\text{circ}(R)$, the total surface density profile within $|z|\leq1.1~\text{kpc}$, $\Sigma(R) \equiv \int_{-1.1\text{kpc}}^{1.1\text{kpc}}\rho_\Phi(R,z)\diff z$, and the ratio of the disk and halo contributions to the total surface density, $\Sigma_\text{disk}(R)/\Sigma_\text{halo}(R)$. We compare the true potential (black lines) with 100 sample potentials (red and blue lines) drawn from the \pdf{} found with MCMC for a \texttt{hot} (red) and a \texttt{cool} (blue) stellar population and also display the relative difference in \% of the true value. (All mock data model parameters are given as Test \ref{test:MW14vsKKS2New} in Table \ref{tbl:tests}.) Overall, the true potential is well recovered---especially in regions where most of the observed stars are located.}
\label{fig:MW14vsKKS2New_contours}
\end{figure*}

\begin{figure*}[!htbp]
\includegraphics[width=1\textwidth]{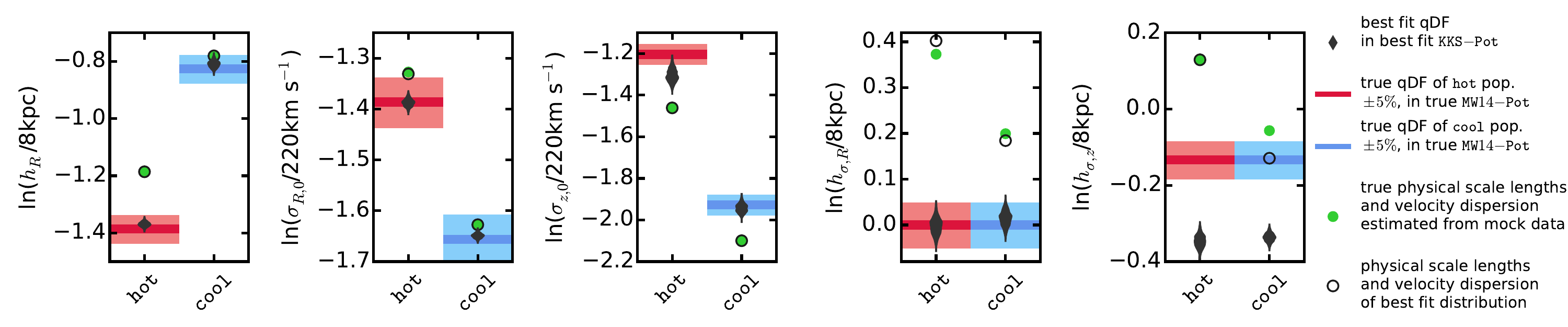}
\caption{Recovery of the qDF parameters for the case where the true and assumed potential deviate from each other (see Test \ref{test:MW14vsKKS2New} in Table \ref{tbl:tests}). The thick red (blue) lines represent the true qDF parameters of the \texttt{hot} (\texttt{cool}) qDF in Table \ref{tbl:referenceMAPs} used to create the mock data, surrounded by a 5\% error region. The grey violins are the marginalized \pdf{}s for the qDF parameters found simultaneously with the potential constraints shown in Figure \ref{fig:MW14vsKKS2New_contours}. We compare the qDF parameters with the actual physical scale lenghts and velocity dispersion at the Sun estimated from the mock data and the best fit distribution by fitting exponential functions to the data. Firstly this shows that---apart from some small deviations in the velocity dispersion scale lengths---the velocity distribution of the mock data is very well reproduced by the best fit. Secondly this demonstrates how the qDF parameters in different potentials do not necessarily agree with each other or with the actual physical velocity distribution.}
\label{fig:MW14vsKKS2New_violins}
\end{figure*}

The comparison between true and best fit potentials are shown in Figure \ref{fig:MW14vsKKS2New_contours}. We find that the potential recovered by \RM{} is in good agreement with the true potential inside of the observed volume of mock tracers. Outside of it we can make predictions at least to a certain extent. Especially the potential forces, to which the stellar orbits are sensitive, are recovered and tightly constrained. This robust recovery of the radial and vertical forces leads to small errors on the estimated circular velocity curve ($\lesssim 5\%$) and surface density within $|z| = 1.1~\text{kpc}$ ($\lesssim 10\%$), respectively. We get the best results for the local density, the surface density and disk-to-halo ratio between $R\sim4~\text{kpc}$ and $R\sim8~\text{kpc}$, i.e., where most of the tracer stars used in the analysis are located (see Figure \ref{fig:MW14vsKKS2New_mockdata_residuals}).

The local density distribution is in general less reliably constrained than the forces, but we still capture the essentials. Exceptions are the inner regions $R\lesssim3~\text{kpc}$, where the \texttt{KKS-Pot} model is missing a bulge by construction, and the local radial density profile, which is somewhat misjudged by the \texttt{KKS-Pot} model. The \texttt{cool} population, where most stars are confined to regions close to the mid-plane, recovers the flatness of the disk better than the \texttt{hot} population, but overall the best fit disk is slightly less dense in the mid-plane than the true disk. While it is in general possible to generate very flattened density distributions from St\"{a}ckel potentials, it might be difficult to simultaneously have a roundish halo and to require that both St\"{a}ckel components have the same focal distance (see Table \ref{tbl:referencepotentials}).

The disk-to-halo surface density fraction within $|z| = 1.1~\text{kpc}$ is not tightly constrained ($\gtrsim20\%$), but recovered within the errors inside of the survey volume. Using a wrong potential model does therefore not necessarily lead to biases in local dark matter measurements.

Overplotted in Figure \ref{fig:MW14vsKKS2New_contours} is also the \texttt{KKS-Pot} with the parameters from Table \ref{tbl:referencepotentials}, which were fixed based on a (by-eye) fit directly to the force field (within $r_\text{max}=4~\text{kpc}$ from the Sun) and rotation curve of the \texttt{MW14-Pot}. The potential found with the \RM{} analysis is an even better fit. This demonstrates that \RM{} fitting infers a potential that in its actual properties resembles the input potential for the mock data in regions of large tracer density as closely as possible, given the differences in functional forms.

Figure \ref{fig:MW14vsKKS2New_violins} compares the true qDF parameters with the best fit qDF parameters belonging to the best fit potentials from Figure \ref{fig:MW14vsKKS2New_contours}, and we also overplot the actual physical scale lengths and velocity dispersion as estimated directly from the mock data. While we recover $h_R$, $\sigma_{R,0}$ and $h_{\sigma,R}$ within the errors, we misjudge the parameters of the vertical velocity dispersion ($\sigma_{0,z}$ and especially $h_{\sigma,z}$), even though the actual mock data distribution is well reproduced. This discrepancy could be connected to the \texttt{KKS-Pot} not being able to reproduce the flatness of the disk. Also, $\sigma_z$ and $\sigma_R$ in Equations \eqref{eq:sigmaRRg}-\eqref{eq:sigmazRg} are scaling profiles for the qDF (cf. BR13) and how close they are to the actual velocity profile depends on the choice of potential; that is, the \emph{physical} velocity dispersion is well recovered, even if the qDF velocity dispersion parameters are not. Figure \ref{fig:MW14vsKKS2New_violins} stresses once more that the actual parameter values of action-based DFs have always to be considered together with the potential in which they were derived. This is of importance in studies that use a fiducial potential to fit action-based DFs to stellar data, like, e.g., \citet{2015MNRAS.449.3479S} and \citet{2016MNRAS.tmp..817D}.

\subsection{The influence of the stellar population's kinematic temperature} \label{sec:results_temperature}

Overall, we found that it does not make a big difference if we use hot or cool stellar populations in our modelling.

How precise and reliable model parameters can be recovered does to a certain extent depend on the kinematic temperature of the data, as well as on the model parameter in question and on the observation volume. But there is no easy rule of thumb, what combination would give the best results (see Figure \ref{fig:isoSph_CLT}). There are two exceptions.

First, the circular velocity at the Sun, $v_\text{circ}(R_\odot)$, is always best recovered with cooler populations (see Figures \ref{fig:isoSphFlexErrConv_SE_vs_error}, \ref{fig:isoSphFlexErrSyst}, \ref{fig:MWdhbMixCont}, \ref{fig:MWdhbMixDiff} and for the recovery of $v_\text{circ}(R)$ at $R \neq R_\odot$ see Figure \ref{fig:MW14vsKKS2New_contours}), because more stars are on near-circular orbits (see Figure \ref{fig:kks2WedgeEx_actions}). Cooler populations are also less sensitive to misjudgements of (spatial) selection functions at large $|z|$ (see Figure \ref{fig:MWdhbIncompR_violins}). There is however the caveat, that cool populations are more susceptible to non-axisymmetric streaming motions in the disk.

Second, hotter populations seem to be less sensitive to misjudgements of proper motion measurement uncertainties (see Figure \ref{fig:isoSphFlexErrSyst}) and pollution with stars from a cooler population (see Figures \ref{fig:MWdhbMixCont} and \ref{fig:MWdhbMixDiff}), because of their higher intrinsic velocity dispersion (see Figure \ref{fig:kks2WedgeEx_xv}).

In addition we find indications in Figure \ref{fig:MW14vsKKS2New_contours}, that different regions within the Galaxy are probed best by populations of different kinematic temperature: The \texttt{hot} population gives the best constraints on the radial local and surface density profiles at a smaller radius than the \texttt{cool} population because of its smaller tracer scale length. The \texttt{cool} population with most stars close to the mid-plane recovers the flatness of the disk more reliably.

\section{Summary and discussion} \label{sec:discussionsummary}

Recently implementations of action DF-based modelling of 6D data in the Galactic disk have been put forth, in part to lay the ground-work for Gaia (BR13; \citealt{,2013MNRAS.433.1411M,2014MNRAS.445.3133P,2015MNRAS.449.3479S}).

We present \RM{}, an improved implementation of the dynamical modelling machinery of BR13, to recover the MW's gravitational potential by fitting an orbit DF to stellar populations within the Galactic disk. In this work we investigated the capabilities, strengths and weaknesses of \RM{} by testing its robustness against the breakdown of some of its assumptions---for well-defined, isolated test cases using mock data. Overall the method works very well and is robust, even when there are small deviations of the model assumptions from the ``real'' Galaxy.

\RM{} applies a full likelihood analysis and is statistically well-behaved. It goes beyond BR13 by allowing for a straightforward and flexible implementation of different model families for potential and DF. It also accounts for selection effects by using full 3D selection functions (given some symmetries).\\

\noindent {\bf Computational speed:~} Large data sets in the age of Gaia require increasingly accurate likelihood evaluations and flexible models. To be able to deal with these computational demands, we sped up the \RM{} code by combining a nested-grid approach with MCMC and by faster action calculation using the St\"{a}ckel \citep{2012MNRAS.426.1324B} interpolation grid by \citet{2015ApJS..216...29B}. Our approach therefore allows us to explore the full \pdf{}, while similar studies (e.g., \citealt{2014MNRAS.445.3133P,2015MNRAS.449.3479S,2016MNRAS.tmp..817D}) focus more on the model with maximum likelihood only. This makes \RM{} also slower: Fitting three \texttt{DHB-Pot} and five qDF parameters in each of the analyses in Tests \ref{test:isoSphFlexIncomp} and \ref{test:MWdhbMix} (see Table \ref{tbl:tests}) takes for example $\sim 25-30$ hours on 25 CPUs. This is still a feasible computational effort as long as we restrict ourselves to potentials with a closed-form expression for $\Phi(R,z)$ (as done in this work). An equivalent analysis using, e.g., a double exponential disk (requiring integrals over Bessel functions) would take several days to weeks for $N_*=20,000$. In any case, the application of \RM{} to millions of stars will be a task for supercomputers and calls for even more improvements and speed-up in the fitting machinery.\\

\noindent {\bf Properties of the data set:~} We could show that \RM{} can provide potential and DF parameter estimates that are very accurate (i.e., unbiased) and precise in the limit of large datasets, as long as the modelling assumptions are fulfilled.

In case the data set is affected by substantive measurement uncertainties, the potential can still be recovered to high precision, as long as these uncertainties are perfectly known and distance uncertainties are negligible. For large proper motion uncertainties, e.g., $\delta \mu \sim 5~\text{mas yr}^{-1}$, the formal errors on the parameters are only twice as large as in the case of no measurement uncertainties. However, properly accounting for measurement uncertainties is computationally expensive.

For the results to be accurate within $2\sigma$ (for 10,000 stars), we need to know to within 10\% both the true stellar distances (at $r_\text{max} \leq 3~\text{kpc}$ and $\delta \mu \lesssim 2 ~ \text{mas yr}^{-1}$) and the true proper motion uncertainties (with $\delta \mu \lesssim 3 ~ \text{mas yr}^{-1}$).

The distance condition is an artefact of the likelihood approximation (Equation \eqref{eq:errorconv}) that \RM{} uses to save computation time, and the reason why we will have to restrict the \RM{} modelling to stars with small distance uncertainties.

Fortunately, the measurement uncertainties of the final Gaia data release with $\delta\mu\lesssim0.3~\text{mas yr}^{-1}$ at $G\lesssim20~\text{mag}$ and $\delta r/r\lesssim5\%$ at $r\sim3~\text{kpc}$ and for $G<15~\text{mag}$ (see Section \ref{sec:results_errors} and \citealt{2014EAS....67...23D}) will be well below these limits and promise accurate potential constraints. Before the final Gaia data release however we might have to restrict the modelling to suitable giant tracers with small uncertainties.

The main caveat of Tests \ref{test:isoSphFlexErrConv_MC_vs_error} and \ref{test:isoSphFlexErrConv_SE_vs_error}-\ref{test:isoSphFlexErrSyst} in Section \ref{sec:results_errors} (see Table \ref{tbl:tests}) concerning measurement uncertainties is the use of the \texttt{Iso-Pot}, which we chose computational speed reasons. However, Tests \ref{test:isoSphFlexIncomp} and \ref{test:MWdhbMix}, which we run for both \texttt{DHB-Pot} and \texttt{Iso-Pot}, gave qualitatively and quantitatively very similar results for both potentials. This makes us confident that also our results about measurement uncertainties are independent of the actual choice of potential.

We also found that the location of the survey volume within the Galaxy matters little. At given sample size a larger survey volume with large coverage in both radial and vertical direction will give the tightest constraints on the model parameters.

The potential recovery with \RM{} seems to be robust against minor misjudgements of the spatial data SF, in particular to a completeness overestimation of $\lesssim 15-20\%$ at the edge of a survey volume with $r_\text{max}=3~\text{kpc}$.

We found indications that populations of different scale lengths and temperature probe different regions of the Galaxy, because the best potential constraints are achieved where most of the stellar tracers are located. This supports the approach by BR13, who measured for each \MAP{} the surface mass density only at one single best radius to account for missing flexibility in their potential model. 

While cooler populations probe the Galaxy rotation curve better and hotter populations are less sensitive to pollution, overall stellar populations of different kinematic temperature seem to be equally well-suited for dynamical modelling.\\

\noindent {\bf Deviations from the DF assumption:~} \RM{} assumes that stellar sub-populations can be described by simple DFs. We investigated how much the modelling would be affected if the assumed family of DFs would differ from the stars' true DF.

In \emph{Example 1} in Section \ref{sec:results_mixedDFs} we considered true stellar DFs being (i) hot with more stars with low velocities and less stars at small radii than assumed (reddish data sets in Figure \ref{fig:MWdhbMix_mockdata_residuals} and \ref{fig:MWdhbMixCont}), or (ii) cool with broader velocity dispersion wings and less stars at large radii than assumed (bluish data sets). We find that case (i) would give more reliable results for the potential parameter recovery.

Binning of stars into \MAPs{} in $[\alpha/\mathrm{Fe}]$ and $[\mathrm{Fe}/\mathrm{H}]$, as done by BR13, could introduce systematic errors due to abundance uncertainties or too large bin sizes---always assuming \MAPs{} follow simple DF families (e.g., the qDF). In \emph{Example 2} in Section \ref{sec:results_mixedDFs} we found that, in the case of 20,000 stars per bin, differences of $\lesssim 20\%$ in the qDF parameters of two neighbouring bins can still give quite good constraints on the potential parameters.

The relative differences in the qDF parameters $\sigma_{R,0}$ and $\sigma_{z,0}$ of neighbouring \MAPs{} in Figure 6 of BR13 (which have bin sizes of $[\mathrm{Fe}/\mathrm{H}] = 0.1$ dex and $\Delta [\alpha/\mathrm{Fe}] = 0.05$ dex) are indeed smaller than $20\%$. For the $h_R$ parameter however the bin sizes in Figure 6 of BR13 might not yet be small enough to ensure no more than $20\%$ of difference in neighbouring bins.

The qDF is a specific example for a simple DF for stellar sub-populations which we used in this paper. But it is not essential for the \RM{} approach. Future studies might apply slight alternatives or completely different DFs to data.\\

\noindent {\bf Gravitational potential beyond the parametrized functions considered:~} In addition to the DF, \RM{} also assumes a parametric model for the gravitational potential. We test how using a potential of St\"{a}ckel form (\texttt{KKS-Pot}, \citealt{1994AA...287...43B}) affects the \RM{} analysis of mock data from a different potential family with halo, bulge and exponential disk (\texttt{MW14-Pot}, \citealt{2015ApJS..216...29B}). The potential recovery is quite successful: We properly reproduce the mock data distribution in configuration space; and the best fit potential is---within the limits of the model---as close as it gets to the true potential, even outside of the observation volume of the stellar tracers. 

For as many as 20,000 stars constraints become already so tight that it should presumably be possible to distinguish between different parametric MW potential models (e.g., the \texttt{DHB-Pot} and the \texttt{KKS-Pot}).

Fitting parametrized potentials of St\"{a}ckel form to MW data (see, e.g., \citealt{1994AA...287...43B,2003MNRAS.340..752F}) has the advantage of allowing action calculations that are accurate and fast. It does however limit the space of potentials that can be investigated, as different potential components are all required to have the same focal distance. Using the \emph{St\"{a}ckel fudge} \citep{2012MNRAS.426.1324B} together with parametrized potentials made up from physically motivated building blocks (exponential disks, power-law dark matter halo etc.), as was done by BR13, seems to be the most promising approach---even though there remain still several challenges concerning computational speed to be solved.\\

\noindent {\bf Different modelling approaches using action-based DFs:~} BR13 focussed on \MAPs{} for a number of reasons: First, they seem to permit simple DFs \citep{2012ApJ...751..131B,2012ApJ...755..115B,2012ApJ...753..148B}, i.e., approximately qDFs \citep{2013MNRAS.434..652T}. Second, all stars must orbit in the same potential. While each \MAP{} can yield different DF parameters, it will also provide a (statistically) independent estimate of the potential. This allows for a valuable cross-checking reference. In some sense, the \RM{} approach focusses on constraining the potential, treating the DF parameters as nuisance parameters. That we were able to show in this work that \RM{} results are quite robust to the form of the DF not being entirely correct motivates this approach further. 

\citet{2014MNRAS.437.2230M} introduced a framework which avoids specific parametrizations of action-based DFs and marginalizes over all possible DFs to constrain the potential. While this is the proper way to treat a nuisance DF, it appears to be computationally very challenging.

For reasons of galaxy and chemical evolution, the DF properties are astrophysically linked between different \MAPs{} \citep{2015MNRAS.449.3479S}. In its current implementation, \RM{} treats all \MAPs{} as independent and does not exploit such correlations. Ultimately, the goal is to do a consistent chemodynamical model that simultaneously fits the potential and $\text{DF}(\vect{J},\text{[X/H]})$ (where $[\mathrm{X}/\mathrm{H}]$ is $[\mathrm{Fe}/\mathrm{H}]$ and other elements either referenced to $\mathrm{H}$ or $\mathrm{Fe}$, i.e., $[\mathrm{X}/\mathrm{H}]$ denotes the whole abundance space) with a full likelihood analysis. This has not yet been attempted with \RM{}, because the behaviour is quite complex. 

Since the first application of \RM{} by BR13 there have been two similar efforts to constrain the Galactic potential and/or orbit DF for the disk:

\citet{2014MNRAS.445.3133P} fitted both potential and a $f(\vect{J})$ to giant stars from the RAVE survey \citep{2006AJ....132.1645S} and the vertical stellar number density profiles in the disk by \citet{2008ApJ...673..864J}. They did not include any chemical abundances in the modelling. Instead, they used a superposition of action-based DFs to describe the overall stellar distribution at once: a superposition of qDFs for cohorts in the thin disk, a single qDF for the thick disk stars and an additional DF for the halo stars. Taking proper care of the selection function requires a full likelihood analysis, which is computationally expensive. \citet{2014MNRAS.445.3133P} choose to circumvent this difficulty by directly fitting (a) histograms of the three velocity components in eight spatial bins to the velocity distribution predicted by the DF and (b) the vertical density profile predicted by the DF to the profiles by \citet{2008ApJ...673..864J}. The vertical force profile of their best fit mass model nicely agrees with the results from BR13 for $R>6.6~\text{kpc}$. The disadvantage of their approach is, that by binning the stars spatially, a lot of information is not used.

\citet{2015MNRAS.449.3479S} have focussed on understanding the abundance-dependence of the DF, relying on a fiducial potential. They developed extended distribution functions (eDF), i.e., functions of both actions and metallicity for a superposition of thin and thick disk, each consisting of several cohorts described by qDFs, a DF for the halo, a functional form of the metallicity of the interstellar medium at the time of birth of the stars, and a simple prescription for radial migration. They applied a full likelihood analysis accounting for selection effects and found a best fit for the eDF in the fixed fiducial potential by \citet{1998MNRAS.294..429D} to the stellar phase-space data of the Geneva-Copenhagen Survey \citep{2004A&A...418..989N,2009A&A...501..941H}, metallicity determinations by \citet{2011A&A...530A.138C} and the stellar density curves by \citet{1983MNRAS.202.1025G}. Their best fit predicted the velocity distribution of SEGUE G-dwarfs \citep{2014ApJS..211...17A} quite well, but had biases in the metallicity distribution, which they accounted to being a problem with the SEGUE metallicities. 

\citet{2016MNRAS.tmp..817D} proceeded recently in a similar fashion to constrain an eDF for halo stars.\\

\noindent {\bf Future work:~} We know that real galaxies, including the MW, are not axisymmetric. Using N-body models, we will explore in a subsequent paper how the recovery of the gravitational potential with \RM{} will be affected when data from a non-axisymmetric disk galaxy system with spiral arms get interpreted through axisymmetric models. There are several interesting scientific questions for which a \RM{} investigation of galaxy simulations could be a pragmatic approach to address them: (i) What is the influence of spiral arms and resonances on the modelling outcome? (ii) Can we recover the potential well enough to calculate actions so accurate that clumps in orbit space can be identified? This is important to be able to compare clumps in action space to clustering of stars in abundance space. (iii) How do results from \RM{}, i.e., the potential and DF, compare with results from Jeans models?


\section{Acknowledgements}

We thank Glenn van de Ven for suggesting the use of Kuzmin-Kutuzov St\"{a}ckel potentials in this case study, the anonymous referee for her/his very helpful comments and suggestions, James J. Binney and Payel Das (University of Oxford) for valuable discussions, and Rene Andrae (MPIA) for an estimate of Gaia distance uncertainties. W.H.T. and H.-W.R. acknowledge funding from the European Research Council under the European Union's Seventh Framework Programme (FP 7) ERC Grant Agreement n. [321035]. J.B. acknowledges the financial support from the Natural Sciences and Engineering Research Council of Canada. 

\bibliography{References_Trick_MW_Modelling}{}
\bibliographystyle{aasjournal}

\clearpage
\LongTables
\begin{landscape}
\begin{deluxetable}{lllll}
\tabletypesize{\scriptsize}
\tablecaption{Summary of test suites in this work: The first column indicates the test suite, the second column the potential, DF and SF model, etc., used for the mock data creation, the third column the corresponding model assumed in the \RM{} analysis, and the last column lists the figures belonging to the test suite and summarizes the results. Reference potentials and qDFs are introduced in Tables \ref{tbl:referencepotentials} and \ref{tbl:referenceMAPs}, respectively. Parameters that are not left free in the analysis, are always fixed to their true value. Unless stated otherwise, all mock data sets have SFs with completeness$(\vect{x}) = 1$ and no measurement uncertainties, and we use $N_x=16, N_v = 24, n_\sigma = 5$ as numerical accuracy for calculating the likelihood normalisation (see Section \ref{sec:likelihood_normalisation}). \label{tbl:tests}}
\tablewidth{0pt}
\tablehead{
\colhead{Test} & & \colhead{Model for Mock Data} & \colhead{Model in Analysis} & \colhead{Figures \& Results}}
\startdata
Test \testlabel{test:norm_accuracy} {1}: & \emph{Potential:} & \texttt{DHB-Pot} & - & Figure \ref{fig:norm_accuracy}\\
Numerical accuracy & \emph{DF:} & \texttt{hot} or \texttt{cool} qDF & & Suitable accuracy for our tests: \\
in calculating & \emph{Survey volume:} & sphere around Sun, $r_\text{max} = 0.2, 1, 2, 3$ or $4~\text{kpc}$ & & $N_x=16, N_v = 24, n_\sigma = 5$.\\
the likelihood & \emph{Numerical accuracy:} & $N_x\in[5,32]$, $N_v\in[4,48]$, $n_\sigma\in[3,7]$ & & Higher spatial resolution is\\
normalisation & & & & required for cooler populations.\\
\tableline
Test \testlabel{test:isoSphFlexErrConv_MC_vs_error}{2} :		& \emph{Potential:} 	& \texttt{Iso-Pot} & \texttt{Iso-Pot}, all parameters free & Figure \ref{fig:isoSphFlexErrConv_MC_vs_error}\\
Numerical convergence 	& \emph{DF:}			& \texttt{hot} qDF & qDF, all parameters free & The number of required MC samples\\
of convolution		& \emph{Survey Volume:}	& sphere around Sun, $r_\text{max} = 3~\text{kpc}$ & (fixed \& known) & scales as $N_\text{samples} \propto (\delta v_\text{max})^2$,\\
with measurement		& \emph{Uncertainties:}		& $\delta \text{RA} =\delta \text{Dec} =\delta(m-M)=0$	& (fixed \& known)	& with $\delta v_\text{max}$ being the largest $\delta v$\\
uncertainties					&						& $\delta v_\text{los} = 2~\text{km s}^{-1}$ & & in the data set. If $N_*$ is smaller,\\
						&						& $\delta \mu =$ 2,3,4 or $5~\text{mas yr}^{-1}$ & & less accuracy, i.e., less $N_\text{samles}$,\\
						& \emph{Numerical accuracy:} & & $N_\text{samples} \in [25,1200]$& is needed to reach a given\\
						& \emph{$N_{*}$:} & 10,000 & & accuracy.\\	 
\tableline
Test \testlabel{test:isoSphFlex} {3.1}: & \emph{Potential:} & \texttt{Iso-Pot} & \texttt{Iso-Pot}, all parameters free & Figure \ref{fig:isoSphFlex_triangleplot}\\
Shape of the model & \emph{DF:} & \texttt{hot} qDF & qDF, all parameters free & The \pdf{} becomes a multivariate\\
parameters' \pdf{} & \emph{Survey Volume:} & sphere around Sun, $r_\text{max} = 2~\text{kpc}$ & (fixed \& known) & Gaussian in the limit of\\
for large data sets & \emph{$N_{*}$:} & 20,000 & & large data. The width of the\\
	& & & & \pdf{} scales as $1/\sqrt{N_*}$.\\

\tableline
Test \testlabel{test:isoSph_CLT} {3.2}: & \emph{Potential:} & \texttt{Iso-Pot} & \texttt{Iso-Pot}, free parameter: $b$ & Figure \ref{fig:isoSph_CLT}\\
Parameter estimates & \emph{DF:} &\texttt{hot} or \texttt{cool} qDF & qDF, free parameters: & \RM{} behaves like an\\
are unbiased; & & & $\ln h_R,\ln\sigma_{z,0},\ln h_{\sigma,z}$ & unbiased maximum likelihood\\
Influence of survey & \emph{Survey volume:} & sphere around Sun, $r_\text{max} = 0.2, 1, 2, 3$ or $4~\text{kpc}$ & (fixed \& known) & estimator. Larger survey volumes\\
volume size & \emph{$N_{*}$:} & 20,000 & & lead to tighter constraints,\\
& & & & even at the same $N_*$.\\

\tableline
Test \testlabel{test:wedgesVol} {4} :		& \emph{Potential:} 	& (i) \texttt{Iso-Pot} or (ii) \texttt{DHB-Pot} & (i) \texttt{Iso-Pot}, all parameters free & Figure \ref{fig:wedgesVol_bias_vs_SE} \\
Influence of 			& 						& 													& (ii) \texttt{DHB-Pot}, free parameters: & The exact position \& shape of the\\
position \& shape 		& 			& 										& $v_\text{circ}(R_\odot)$, $a_\text{disk}$, $f_\text{halo}$ & survey volume plays only a\\
of survey volume 		& \emph{DF:} & \texttt{hot} qDF											& qDF, all parameters free & minor role. Having both large\\
on parameter recovery 	& \emph{Survey volume:}	& 4 different wedges, see Figure \ref{fig:wedgesVol_bias_vs_SE}, upper panel & (fixed \& known) & radial and vertical extent should\\
						& \emph{$N_{*}$:} & 20,000 & & give the tightest constraints.\\
\tableline
Test \testlabel{test:isoSphFlexIncomp}{5} : & \emph{Potential:} & \texttt{DHB-Pot} & \texttt{DHB-Pot}, free parameters: & Figures \ref{fig:MWdhbIncompR_violins} \\
Influence of & & & $v_\text{circ}(R_\odot), a_\text{disk}, f_\text{halo}$ & For minor misjudgements\\
wrong assumptions & \emph{DF:} & \texttt{hot} or \texttt{cool} qDF & qDF, all parameters free & of a radially symmetric SF\\
about the spatial SF & \emph{Survey volume:} & sphere around Sun, $r_\text{max} = 3~\text{kpc}$ & (fixed \& known) & (i.e., $\epsilon_r \lesssim 0.15$ for \texttt{hot} and\\
on parameter recovery & \emph{Completeness:} & Equation \eqref{eq:rad_incomp} & completeness$(\vect{x})$ = 1, & $\epsilon_r \lesssim 0.2$ for \texttt{cool} populations)\\
 & & with $\epsilon_r \in [0,0.7]$ & i.e., $\epsilon_r=0$& the potential recovery\\
 & \emph{$N_{*}$:} & 20,000 & & is still robust.\\
\tableline
Test \testlabel{test:isoSphFlexErrConv_SE_vs_error}{6.1} : & \emph{Potential:} & \texttt{Iso-Pot} & \texttt{Iso-Pot}, all parameters free & Figure \ref{fig:isoSphFlexErrConv_SE_vs_error}\\
Effect of proper motion & \emph{DF:} & \texttt{hot} or \texttt{cool} qDF & qDF, all parameters free & If $\delta \mu$ is perfectly known, the\\
uncertainties on & \emph{Survey volume:} & sphere around Sun, $r_\text{max} = 3~\text{kpc}$ & (fixed \& known) & precision of the potential para-\\
precision of & \emph{Uncertainties:} & (i) $\delta\text{RA}=\delta\text{Dec}=\delta(m-M)=0,$ & (fixed \& known) & meter recovery is only a\\
potential recovery & & $\delta v_\text{los} = 2~\text{km s}^{-1},$ & & factor $\sim1.15-2$ worse\\
& & $\delta \mu = 1, 2, 3, 4$ or $5~\text{mas yr}^{-1}$ & & for $\delta\mu\sim1-5~\text{mas yr}^{-1}$\\
& & (ii) no measurement uncertainties & & than for a data set without\\
 & \emph{$N_{*}$:} & 10,000 & & proper motion uncertainties.\\
\\
Test \testlabel{test:isoSphFlexErrConv_bias_vs_SE}{6.2} : & \emph{Potential:} 	& \texttt{Iso-Pot} & \texttt{Iso-Pot}, all parameters free & Figure \ref{fig:isoSphFlexErrConv_bias_vs_SE}\\
Testing the convolution		& \emph{DF:}			& \texttt{hot} qDF & qDF, all parameters free & The approximate likelihood\\
with measurement 		& \emph{Survey Volume:}	& sphere around Sun, $r_\text{max} = 3~\text{kpc}$ & (fixed \& known) & in Equation \eqref{eq:errorconv} is the true\\
 uncertainties in Equation \eqref{eq:errorconv} & \emph{Uncertainties:}		& $\delta \text{RA} =\delta \text{Dec} =0$,	& (fixed \& known)	& likelihood in the absence of\\
with \& without			&						& $\delta v_\text{los} = 2~\text{km s}^{-1}$, & & position errors. For distance\\
distance uncertainties						&						& $\delta \mu =$ 1,2,3,4 or $5~\text{mas yr}^{-1}$, & & uncertainties $\lesssim10\%$ at\\
						&						& (i) $\delta(m-M) = 0$ or & & $r_\text{max}\sim3~\text{kpc}$, which is the case\\
						& & (ii) $\delta(m-M) \neq 0$ (see Figure \ref{fig:isoSphFlexErrConv_bias_vs_SE}) & & for Gaia and $G \lesssim 15$, the intro-\\
						& \emph{$N_{*}$:} & 10,000 & & duced bias is still less than $2\sigma$.\\
\tableline
Test \testlabel{test:isoSphFlexErrSyst}{6.3} :	& \emph{Potential:}		& \texttt{Iso-Pot} & \texttt{Iso-Pot}, all parameters free & Figure \ref{fig:isoSphFlexErrSyst}\\
Underestimation 	& \emph{DF:}			& \texttt{hot} or \texttt{cool} qDF & qDF, all parameters free & The bias introduced by under-\\
of proper motion 	& \emph{Survey volume:}	& sphere around Sun, $r_\text{max} = 3~\text{kpc}$ & (fixed \& known) & estimating $\delta\mu$ grows with $\delta \mu$.\\
uncertainties 			 	& \emph{Uncertainties:}		& only proper motion uncertainties & proper motion uncertainties & If $\delta \mu \lesssim 3~\text{mas yr}^{-1}$ and $\delta \mu$\\
					&						& 1, 2 or $3~\text{mas yr}^{-1}$ & 10\% or 50\% underestimated & is underestimated by 10\%\\
					& \emph{$N_{*}$:} & 10,000 & & the bias is only $\sim2\sigma$ or less.\\
\tableline
Test \testlabel{test:MWdhbMix}{7} : & \emph{Potential:} & \texttt{DHB-Pot} & \texttt{DHB-Pot}, free parameters: & Figures \ref{fig:MWdhbMix_mockdata_residuals}, \ref{fig:MWbdhMix_residual_RvsZ}, \ref{fig:MWdhbMixCont} \& \ref{fig:MWdhbMixDiff}\\
Deviations of the & & & $v_\text{circ}(R_\odot), a_\text{disk}, f_\text{halo}$ & \emph{Example 1:}\\
assumed DF & \emph{DF:} & mix of two qDFs... & single qDF, all parameters free & Hot qDF-like populations polluted\\
from the & & (i) \emph{Example 1:} & & by up to $\sim 30\%$ of stars from a much\\
stars' true DF & & ... with different mixing rates & & cooler population give still reliable\\
 & & and fixed qDF parameters & & potential constraints. This is not\\
 & & (\texttt{hot} \& \texttt{cooler} qDF from Table \ref{tbl:referenceMAPs}) & & true for the opposite case.\\
 & & (ii) \emph{Example 2:} & & \emph{Example 2:}\\
 & & ... with 50/50 mixing rate & & Differences of $\lesssim 20\%$ in qDF\\
 & & and varying qDF parameters (by $X\%$): & & parameters do not matter\\
 & & \emph{a)} \texttt{hot} \& \texttt{colder} qDF or & & when mixing sub-populations,\\
 &					 & \emph{b)} \texttt{cool} \& \texttt{warmer} qDF (see Table \ref{tbl:referenceMAPs}) & & e.g., due to finite binning in\\
 & \emph{Survey volume:}& sphere around Sun, $r_\text{max}=2~\text{kpc}$ & (fixed \& known) & abundance space and abundance\\
 & \emph{$N_{*}$:} & 20,000 & & errors.\\

\tableline
Test \testlabel{test:MW14vsKKS2New}{8} :			& \emph{Potential:} & \texttt{MW14-Pot} & \texttt{KKS-Pot}, all parameters free, & Figures \ref{fig:MW14vsKKS2New_mockdata_residuals}, \ref{fig:MW14vsKKS2New_contours} \& \ref{fig:MW14vsKKS2New_violins}\\
Deviations of the		& & & only $v_\text{circ}(R_\odot)=230~\text{km s}^{-1}$ fixed & We find a good approximation\\
assumed potential model	& \emph{DF:} & \texttt{hot} or \texttt{cool} qDF & qDF, all parameters free & for the true potential, especially\\
from the stars'			& \emph{Survey volume:} & sphere around Sun, $r_\text{max} = 4~\text{kpc}$ & (fixed \& known) & where most stars are located,\\
true potential & \emph{$N_{*}$:} & 20,000 & & given the limitations of the wrong\\
 & \emph{Numerical accuracy:} & & $N_x=20, N_v = 28, n_\sigma = 5.5$ & assumed potential model family.
\enddata
\end{deluxetable}
\clearpage
\end{landscape}

\appendix
\section{Mock data} \label{app:mockdata}

The mock data in this work is generated according to the following procedure:

We assume that the positions and velocities of our stellar mock sample are indeed drawn from our assumed family of potentials (Section \ref{sec:potentials}) and DFs (Section \ref{sec:qDF}), with given parameters $p_\Phi$ and $p_\text{DF}$. The DF is in terms of actions, while the transformation $(\vect{x}_i,\vect{v}_i) \stackrel{\Phi}{\longrightarrow} \vect{J}_i$ is computationally much less expensive than its inversion. We therefore employ the following efficient two-step method for creating mock data, which also accounts for a spatial survey selection function SF$(\vect{x})$ see Appendix \ref{app:selectionfunction}.

In the first step we draw stellar positions $\vect{x}_i$. We start by setting up the interpolation grid for the tracer density $\rho(R,|z| \mid p_\Phi, p_\text{DF})$ generated according to Section \ref{sec:qDF}.\footnote{For the creation of the mock data we use $N_x = 20$, $N_v = 40$ and $n_\sigma=5$ in Equation \eqref{eq:tracerdensity}.} Next, we sample random positions $(R_i,z_i,\phi_i)$ uniformly within the observable volume. Using a Monte Carlo rejection method we then shape the samples distribution to follow $\rho(R,|z| \mid p_\Phi, p_\text{DF})$. To apply a non-uniform completeness function, we use the rejection method a second time. The resulting set of positions $\vect{x}_i$ follows the distribution $p(\vect{x}) \propto \rho_\text{DF}(R,|z| \mid p_{\Phi},p_\text{DF}) \times \text{SF}(\vect{x})$.

In the second step we draw velocities $\vect{v}_i$. For each of the positions $(R_i,z_i)$ we first sample velocities from a Gaussian envelope function in velocity space which is then shaped towards DF$(\vect{J}[R_i,z_i,\vect{v} \mid p_{\Phi}] \mid p_\text{DF})$ using a rejection method. We now have a mock data set satisfying $(\vect{x}_i,\vect{v}_i) \longrightarrow p(\vect{x},\vect{v}) \propto \text{DF}(\vect{J}[\vect{x},\vect{v} \mid p_{\Phi}] \mid p_\text{DF}) \times \text{SF}(\vect{x})$.

Measurement uncertainties can be added to the mock data by applying the following modifications to the above procedure. We assume Gaussian uncertainties in the heliocentric phase-space coordinates $\tilde{\vect{x}} = (\text{RA},\text{Dec},(m-M)), \tilde{\vect{v}} = (\mu_\text{RA} \cdot \cos \text{Dec} ,\mu_\text{Dec},v_\text{los})$ (see Section \ref{sec:coordinates}). In the case of distance and position uncertainties stars virtually scatter in and out of the observed volume. To account for this, we draw the \emph{true} $\vect{x}_i$ from a volume that is larger than the actual observation volume, perturb the $\vect{x}_i$ according to the position uncertainties and then reject all stars that lie now outside of the observed volume. This mirrors the random scatter around the detection threshold for stars whose distances are determined from the apparent brightness and the distance modulus. We then sample \emph{true} $\vect{v}_i$ (given the \emph{true} $\vect{x}_i$) as described above and perturb them according to the velocity uncertainties.

\begin{figure}[!htbp]
\centering
\begin{minipage}{0.48\textwidth}
\centering
\includegraphics[width=\textwidth]{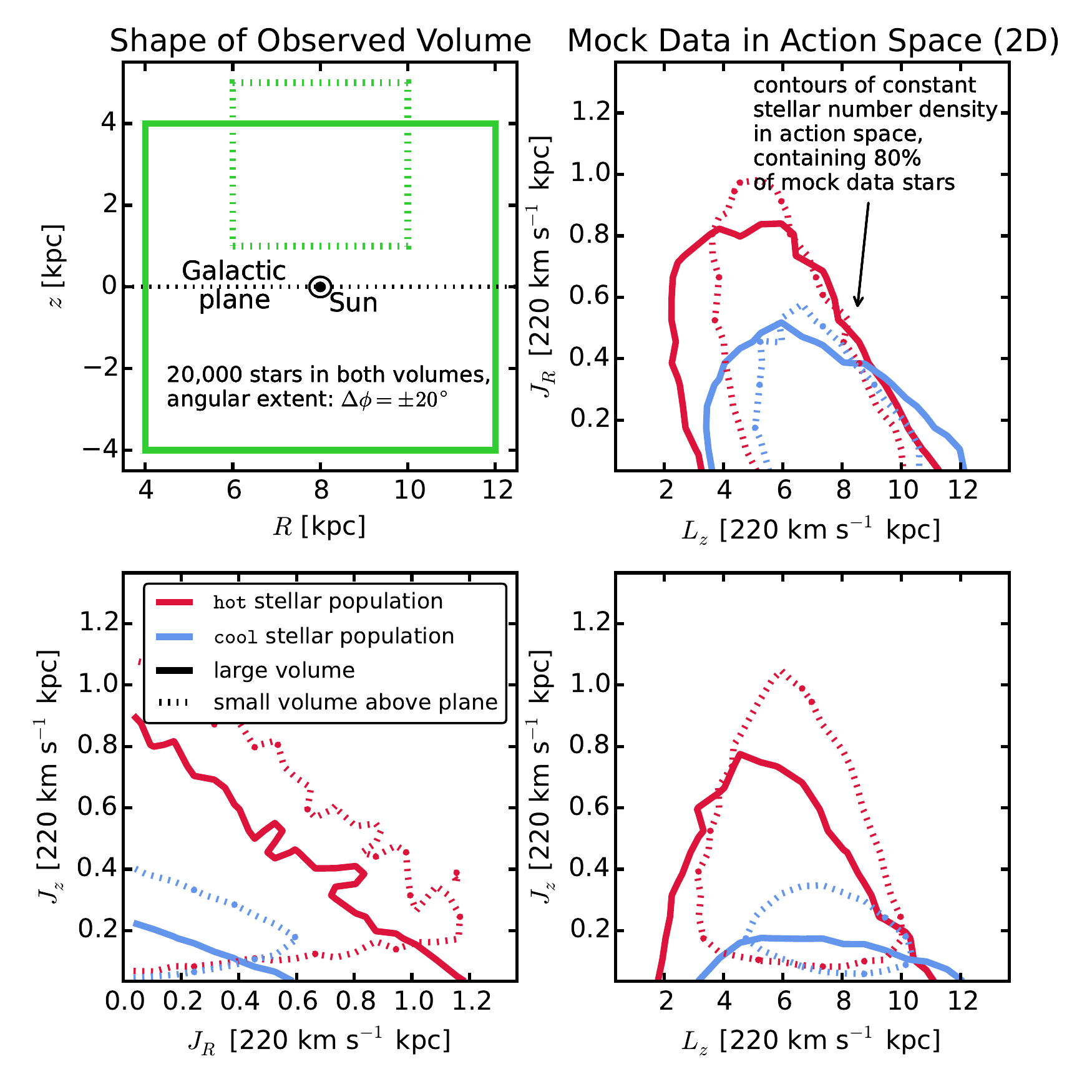}
\caption{Distribution of mock data in action space (2D iso-density contours, enclosing 80\% of the stars), depending on shape and position of a wedge-like survey observation volume (upper left panel, see also Appendix \ref{app:selectionfunction}) and temperature of the stellar population (indicated in the legend). The four mock data sets are generated in the \texttt{KKS-Pot} from Table \ref{tbl:referencepotentials} from either the \texttt{hot} or \texttt{cool} DF in Table \ref{tbl:referenceMAPs}. The distribution in action space visualizes how orbits with different actions reach into different regions within the Galaxy. The corresponding mock data in configuration space is shown in Figure \ref{fig:kks2WedgeEx_xv}.} 
\label{fig:kks2WedgeEx_actions}
\end{minipage}
\hfill
\begin{minipage}{0.48\textwidth}
\centering
\includegraphics[width=\textwidth]{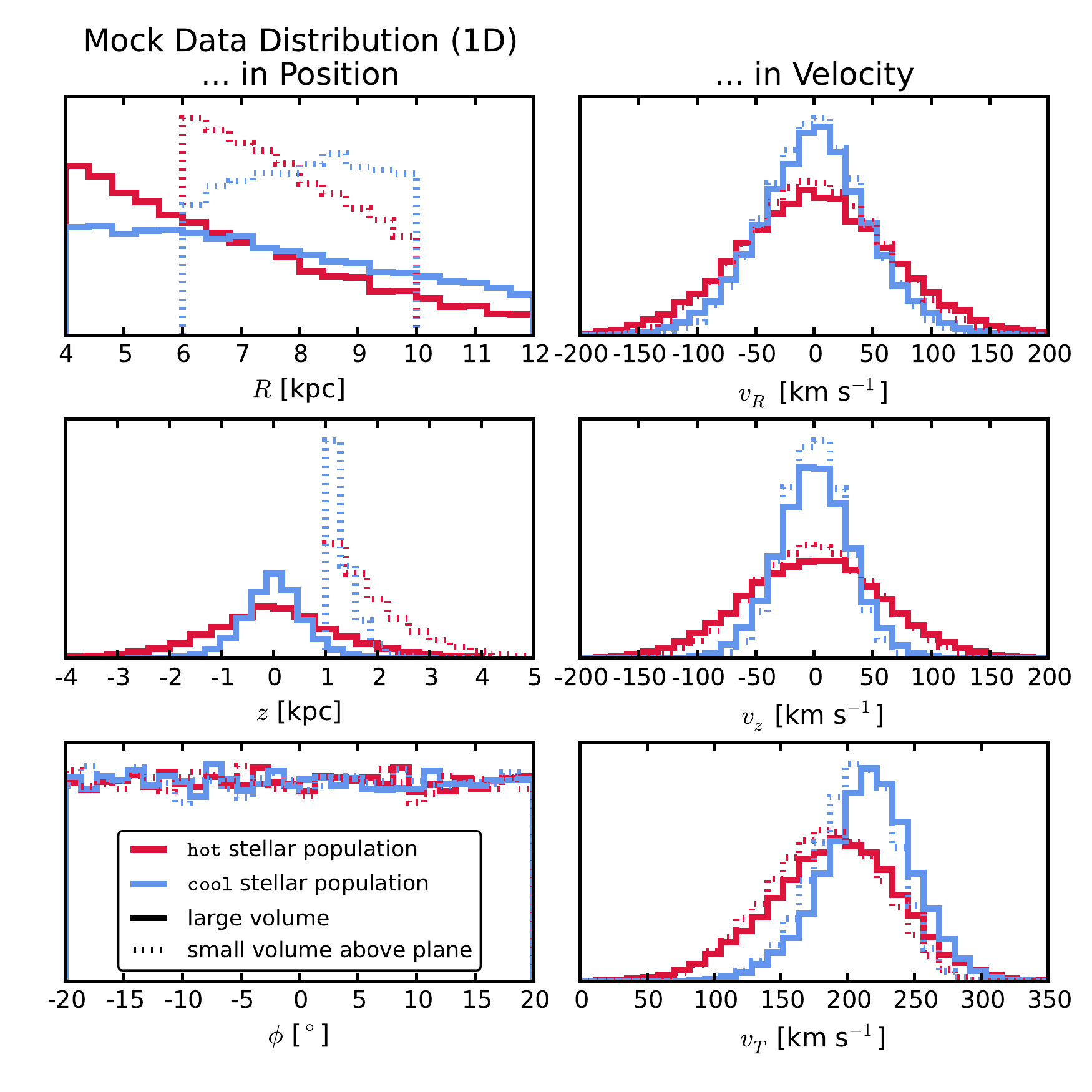}
\caption{Distribution of the mock data from Figure \ref{fig:kks2WedgeEx_actions} in configuration space. The corresponding observation volumes (as indicated in the legend) are shown in Figure \ref{fig:kks2WedgeEx_actions}, upper left panel. The 1D histograms illustrate that qDFs generate realistic stellar distributions in Galactocentric coordinates $(R,z,\phi,v_R,v_z,vT)$: More stars are found at smaller $R$ and $|z|$, and are distributed uniformly in $\phi$ according to our assumption of axisymmetry. The distribution in radial and vertical velocities, $v_R$ and $v_z$, is approximately Gaussian with the (total projected) velocity dispersion being of the order of $\sim\sigma_{R,0}$ and $\sim\sigma_{z,0}$ (see Table \ref{tbl:referenceMAPs}). The distribution of tangential velocities $v_T$ is skewed because of asymmetric drift.} 
\label{fig:kks2WedgeEx_xv}
\end{minipage}
\end{figure}

We show examples of mock data sets (without measurement uncertainties) in action space (Figure \ref{fig:kks2WedgeEx_actions}) and configuration space $(\vect{x},\vect{v})$ (Figure \ref{fig:kks2WedgeEx_xv}). The mock data generated from the qDF follow the expected distributions in configuration space. The distribution in action space illustrates the intuitive physical meaning of actions: The stars of the \texttt{cool} population have in general lower radial and vertical actions, as they are on more circular orbits. Circular orbits with $J_R = 0$ and $J_z = 0$ can only be observed in the Galactic mid-plane. The different ranges of angular momentum $L_z$ in the two example observation volumes reflect $L_z \sim R \times v_\text{circ}$ and the volumes' different radial extent. The volume at larger $z$ contains stars with higher $J_z$. An orbit with $L_z \ll$ or $\gg L_z(R_\odot)$ can only reach into a volume at $\sim R_\odot$, if it is more eccentric and has therefore larger $J_R$. This together with the effect of asymmetric drift explains the asymmetric distribution of $J_R$ vs. $L_z$ in Figure \ref{fig:kks2WedgeEx_actions}.

\section{Selection functions} \label{app:selectionfunction}

Any survey's selection function (SF) can be understood as defining an effective sample sub-volume in the space of observables, e.g., position on the sky (limited by the pointing of the survey), distance from the Sun (limited by brightness and detector sensitivity), colors and metallicity of the stars (limited by survey mode and targeting). The SF can therefore be thought of as having both spatial small scale structure (due to pencil beam pointing, dust obscuration, etc.) and some overall spatial characteristics (e.g., mean height above the plane and mean Galactocentric radius of the stars). The treatment of realistic and complex SFs was already demonstrated in BR13 (who used the pencil-beam SF of the SEGUE survey \citep{2012ApJ...753..148B}) and \citet{2016ApJ...818..130B} (who investigated the effect of dust extinction). In this work we aim to make a generic and basic exploration of search volume shapes and, as shown by \citet{2016ApJ...818..130B}, this should be possible without explicitly considering spatial SF substructure. Inspired by the contiguous nature of the Gaia SF, which is basically only limited by a magnitude cut, and the fact that this magnitude cut would---in the absence of small scale structure---translate to a sharp distance cut for standard candle tracer populations like red clump stars, we therefore use in our modelling a simple spatial SF of spherical shape with radius $r_\text{max}$ around the Sun,
\begin{eqnarray}
\text{SF}(\vect{x}) \equiv \begin{cases}
\text{completeness}(\vect{x}) &\text{if $|\vect{x}-\vect{x}_\odot| \leq r_\text{max}$,}\\
0 & \text{otherwise,}
\end{cases} \label{eq:selectionfunction}
\end{eqnarray}

We set $0 \leq \text{completeness}(\vect{x}) \leq 1$ everywhere inside the observed volume, so it can be understood as a position-dependent detection probability for a star at $\vect{x}$. Unless explicitly stated otherwise, we simplify to $\text{completeness}(\vect{x}) = 1$. Additionally, we use in Figure \ref{fig:wedgesVol_bias_vs_SE} (Test \ref{test:wedgesVol}) and Figures \ref{fig:kks2WedgeEx_actions}-\ref{fig:kks2WedgeEx_xv} for illustrative purposes some rather unrealistic survey volumes which are angular segments of a cylindrical annulus (wedge), i.e., the volume with $R \in [R_\text{min},R_\text{max}],\phi \in [\phi_\text{min},\phi_\text{max}],z \in [z_\text{min},z_\text{max}]$ within the model Galaxy. 


\end{document}